\documentstyle[emulateapj,psfig]{article}
\input epsf
 
\begin{document}
\submitted{Recieved 1999 November 28th, accepted 2000 September 18th}

\title{A Serendipitous Galaxy Cluster Survey with {\em XMM\/}: Expected Catalogue Properties and Scientific Applications}
\author{ A.~Kathy Romer}
\affil{Physics Department, Carnegie Mellon University, Pittsburgh, 
PA15213; romer@cmu.edu}
\author{Pedro T.~P.~Viana\altaffilmark{1}}
\affil{Centro de Astrof\'{\i}sica da Universidade do Porto,
Rua das Estrelas s/n, 4150 Porto, Portugal}
\altaffiltext{1}{Also affiliated with: 
Departamento de Matem\'{a}tica Aplicada da Faculdade de Ci\^{e}ncias 
da Universidade do Porto.}
\author{Andrew R.~Liddle\altaffilmark{2}}
\affil{Astrophysics Group, Imperial College, Blackett Laboratory,   Prince Consort Road, London SW7 2BZ, UK}
\altaffiltext{2}{Current address: 
Astronomy Centre, University of Sussex, Brighton BN1 9QJ, UK.}
\author{Robert G.~Mann\altaffilmark{3}}  
\affil{Astrophysics Group, Imperial College, Blackett Laboratory,   Prince Consort Road, London SW7 2BZ, UK}
\altaffiltext{3}{Current address: Institute for Astronomy, University of Edinburgh, Royal Observatory, Blackford Hill, Edinburgh, EH9 3NJ, UK.}

\begin{abstract}

This paper describes a serendipitous galaxy cluster survey that we
plan to conduct with the {\em XMM\/} X--ray satellite.  We have
modeled the expected properties of such a survey for three different
cosmological models, using an extended Press--Schechter (Press \&
Schechter~1974) formalism, combined with a detailed characterization
of the expected capabilities of the EPIC camera on board {\em XMM\/}.
We estimate that, over the ten year design lifetime of {\em XMM}, the
EPIC camera will image a total of $\simeq800$ square degrees in fields
suitable for the serendipitous detection of clusters of galaxies. For
the presently--favored low-density model with a cosmological constant,
our simulations predict that this survey area would yield a catalogue
of more than 8000 clusters, ranging from poor to very rich systems,
with around 750 detections above $z=1$.  A low-density open Universe
yields similar numbers, though with a different redshift distribution,
while a critical-density Universe gives considerably fewer
clusters. This dependence of catalogue properties on cosmology means
that the proposed survey will place strong constraints on the values
of $\Omega_0$ and $\Omega_\Lambda$.  The survey would also facilitate
a variety of follow-up projects, including the quantification of
evolution in the cluster X-ray luminosity--temperature relation, the
study of high-redshift galaxies via gravitational lensing, follow-up
observations of the Sunyaev-Zel'dovich effect and foreground analyses
of cosmic microwave background maps.

\end{abstract}


\section{Introduction}
\label{intro}

Galaxy clusters are the largest gravitationally-bound structures in
the Universe today and they are proving to be extremely powerful
cosmological probes. In the hierarchical gravitational instability
picture of structure formation, massive clusters arise from the
extreme tail in the distribution of density fluctuations, so their
number density depends critically on the cosmological parameters that
determine the initial rms width, and the evolution with redshift, of
that distribution. It thus follows that the observed cluster number density 
can provide strong constraints on those parameters.  For example, the
number density of clusters at $z=0$ currently offers the most reliable
constraint (Evrard 1989; White, Efstathiou \& Frenk 1993) on the
amplitude of density perturbations on small scales, as quantified by
$\sigma_8$ -- the rms mass fluctuation in spheres of radius 8$h^{-1}$
Mpc, where $h$ is the Hubble constant, $H_0$, in units of 100
km$^{-1}$ Mpc$^{-1}$. In addition, several authors (White et
al.~1993b; Bludman 1998; Gheller, Pantano \& Moscardini 1998; Arnaud
\& Evrard 1999; Wu \& Xue 1999) have tried to estimate the ratio of
baryonic to non-baryonic matter in the Universe as a whole from the
observed baryon fraction in clusters. Perhaps the most exciting
prospect is the possibility (Oukbir \& Blanchard~1992; Viana \& Liddle
1996,~1999) of constraining the matter density, $\Omega_0$, (and,
perhaps, $\Omega_\Lambda\equiv \Lambda/3H_0^2$, where $\Lambda$ is the
cosmological constant) by observing the evolution of the number
density of rich clusters. A great deal of attention (Henry 1997;
Bahcall \& Fan 1998; Eke et al.~1998; Sadat, Blanchard \& Oukbir 1998;
Blanchard et al.~1999; Borgani et al.~1999; Reichart et al.~1999a;
Viana \& Liddle 1999) has been paid in recent years to this issue.  To
date, no consensus as to the value of $\Omega_0$ has been reached,
due, in large part, to the inadequacies of the cluster catalogues
currently available.

The inadequacies of current cluster catalogues motivates the creation
a major new galaxy cluster catalogue using ESA's X-ray Multi-Mirror
({\em XMM\/}) satellite.  The {\em XMM} satellite\footnote{XMM Home
Page: {\tt astro.estec.esa.nl/XMM/}} was successfully launched on
December 10th 1999. It is a multi--mirror instrument, comprising of
three Wolter type-1 X-ray telescope modules. There is an EPIC
(European Photon Imaging Camera) imaging detector in the focal plane
of each of the three telescope modules. The field of view of two of
the EPIC detectors is paved with 7 MOS CCDs, while the third is paved
with 12 pn CCDs. The MOS detectors share the focal plane of their
respective telescope modules with an RGS (Reflection Grating
Spectrometer) camera.  All five detectors work simultaneously, meaning
that every {\em XMM} pointed observation will yield the type of
imaging data required for serendipitous source detection.  (This is in
contrast to {\em Chandra\/}, which allows for either imaging or
grating observations, but not both at the same time.)

To illustrate the enhanced sensitivity of {\em XMM} over other X-ray
satellites, we have calculated, using the {\tt fakeit} and {\tt show
rates} commands in {\sc xspec} (version 10.00, Arnaud 1996), the {\em
XMM}, {\em Chandra\/}, {\em ROSAT\/} and {\em Einstein\/} count rates
for an absorbed Raymond--Smith (Raymond \& Smith 1977) spectrum.
The Raymond--Smith model has 4 input parameters; electron temperature
($T$), metallicity ($Z$), redshift ($z$) and normalization.  For this
comparison, we chose $T=1$ keV, $Z=0.3Z_{\odot}$, $z=0.1$ and set the
normalization so that the model spectrum had an unabsorbed flux of
$1\times10^{-13}$ erg s$^{-1}$ cm$^{-2}$ in the 0.5-2.0 keV band.  
Photo-electric
absorption (with $n_H=4\times10^{20}$ cm$^{-2}$) was included via the
{\sc xspec} {\tt wabs} model, which is based on cross sections
presented in Morrison \& McCammon (1983). The resulting count rates
are 
\begin{tabbing} 
~~~ \= {\em EPIC--pn} (thin filter)~~~~ \= 0.5--10 keV~~~~~~ \= 0.078 s$^{-1}$\\
\> {\em Chandra--Acis-I} \> 0.5--10 keV\> 0.017 s$^{-1}$\\ 
\> {\em ROSAT--PSPC} \> 0.5--2.0 keV \> 0.0088 s$^{-1}$\\ 
\> {\em Einstein--IPC} \> 0.3--3.5 keV \> 0.0040 s$^{-1}$
\end{tabbing} 
This exercise demonstrates that 
{\em XMM} is $\sim 4$ times more sensitive than {\em Chandra},
$\sim 10$ times more sensitive than {\em ROSAT} and $\sim 20$ times
more sensitive than {\em Einstein}. (The response matrices used for
these calculations were epn\_new\_rmf.fits \& epn\_thin\_arf.fits for
{\em XMM}\footnote{Available from {\tt astro.estec.esa.nl} in the
directory tree {\tt /pub/XMM/EPIC/March99/RESPONSES}},
w215c2r\_norm.rmf \& w215c2r\_norm.arf for {\em
Chandra}\footnote{Available from {\tt http://asc.harvard.edu}},
pspcb\_gain2\_256.rmf for {\em ROSAT\/}\footnote{Available from {\tt
http://heasarc.gsfc.nasa.gov/}} and ipc\_90jun07\_16ch.rsp for {\em
Einstein\/}$^{7}$.)


The high sensitivity of {\em XMM}, combined with its wide field of
view, excellent spatial resolution and spectral coverage, make it
ideal for cluster detection out to redshifts of $z=1$ and beyond. In
this paper, we detail how an {\em XMM} cluster catalogue may be
constructed through serendipitous detections in archival data.  By
examining the many thousands of pointing observations which will be
made with {\em XMM\/}, it will be possible to build up a large sample
of clusters which extends to $z\gtrsim 1$.
In this paper, we make predictions for the numbers and types of
clusters we hope to detect in the proposed {\em XMM\/} cluster survey
(hereafter {\em XCS}), and discuss the impact of the resulting cluster
catalogue on cosmology. We estimate that the {\em XCS\/} will cover
$\sim800$ square degrees (\S\ref{extdist}) to an effective flux limit
of $\sim 1.5 \times 10^{-14}$ erg s$^{-1}$ cm$^{-2}$
 and contain (if $\Omega_0=0.3$) more than 8000 clusters 
(\S\ref{results:catalogues}).

In \S\ref{other_surveys} we compare the {\em XCS} to existing
and proposed cluster surveys. In \S\ref{ps_model} we construct 
a theoretical model for the cluster population, based on the extended 
Press--Schechter (Press \& Schechter 1974) formalism of Viana \& Liddle (1999).
In \S\ref{assumptions}, we describe the various assumptions we have 
made about the instrument response and about the spatial and spectral
properties of the clusters to be observed. In \S\ref{results} we describe
how we estimated the sensitivity limits of the {\em XCS} and how we
went on to use those limits, in combination with the results of
\S\ref{ps_model}, to produce simulated cluster catalogues. 
Finally, in \S\ref{discussion}, we describe some of the potential
scientific applications of the {\em XCS\/} and discuss
some of the limitations of our calculations. 

Throughout this paper we assume $h=0.5$. 
{\em XMM\/} count rates are quoted in the 0.5--10 keV band and, except
where stated, fluxes and luminosities are quoted in the 0.5--2.0 keV 
band pass.

\section{Cluster Surveys}\label{other_surveys}

Cluster catalogues have traditionally been constructed  
by identifying enhancements in the surface density
of optical galaxies on the sky (e.g~Abell 1958;
Abell et al.~1989) . While this can be made objective and
algorithmic (Lumsden et al.~ 1992; Dalton et al.~1992; Postman et
al.~1996), the projection effects that plague this approach cannot be
overcome completely (van Haarlem et al.~1997). The small angular size
of the X-ray emitting region in a cluster core, and its high contrast
against the background X-ray sky, makes X-ray observations one of 
the best strategies for cluster detection.
 
At low redshift, attention has focussed on the {\em ROSAT\/} All--Sky
Survey ({\em RASS\/}), with a number of cluster samples (Romer et
al.~1994; Ebeling et al.~1996; Henry et al.~1997; Ebeling et al.~1998;
B\"ohringer et al.~ 1998; De Grandi et al.~1999a,b) making use of its
wide areal coverage. The most ambitious of the {\em RASS\/} surveys is
the {\em REFLEX\/} survey, which covers 8235 deg$^2$ (compared to the
$\simeq800$ deg$^2$ covered by the {\em XCS}, see \S\ref{extdist}).
To date only a preliminary sample of {\em REFLEX\/} clusters has been
published (De Grandi et al.~1999b); this sample has a flux limit of
$\sim 4 \times 10^{-12}$ erg s$^{-1}$ cm$^{-2}$, includes 130 clusters
and has a maximum redshift of $z=0.308$. A much larger sample of $\sim
800$ clusters, with a flux limit of $2 \times 10^{-12}$ erg s$^{-1}$
cm$^{-2}$, will be released soon (B\"ohringer et al.~1998). From
Table~\ref{cluster-numbers} we can see that XCS will detect a similar 
total number of clusters at $z < 0.3$, but
this will be in a smaller area and to a deeper flux limit. Further,
essentially all these clusters will be accompanied with serendipitous
temperature measurements (see \S\ref{results:catalogues}). 
By comparison, the largest complete
sample of low--redshift cluster temperatures currently available
contains only 50 objects (Blanchard et al.~1999).

At higher redshifts, the {\em XCS\/} will be far superior to the {\em
RASS\/}--based surveys, since -- with the exception of the {\em NEP}
survey -- the {\em RASS\/}--based surveys do not have the sensitivity
to detect clusters beyond $z\gtrsim 0.3$. The {\em NEP} (North Ecliptic
Pole) survey has higher sensitivity ($\sim1 \times 10^{-13}$ erg
s$^{-1}$ cm$^{-2}$: Gioia 1998) than {\em REFLEX}, due to the scanning
strategy of {\em ROSAT\/}, and has yielded detections of clusters as
distant as $z=0.81$ (Henry et al.~1997). Despite this enhanced
sensitivity, the {\em NEP} survey cannot compete with the {\em XCS\/}, since
the {\em XCS} will cover roughly 10 times the area (800 deg$^2$ compared
to 84.7 deg$^2$) to roughly 10 times the depth. 


Data deeper than the {\em RASS\/} are, therefore, required to detect
high-redshift clusters in significant numbers, and several surveys
have sought them through serendipitous detections in the fields
surrounding {\em Einstein\/} and {\em ROSAT\/} targets. The {\em
Einstein\/} Medium Sensitivity Survey ({\em EMSS\/})
 has the largest areal coverage of any of these (734 deg$^2$
above a 0.3-3.5 keV flux limit of $3.57 \times 10^{-12}$ erg s$^{-1}$
cm$^{-2}$, falling to 40 deg$^2$ above $1.33 \times 10^{-13}$ erg
s$^{-1}$ cm$^{-2}$) yielding a total of 104 clusters (Gioia et
al.~1990; Henry et al.~1992). The measurement
of temperatures for several of these clusters with redshifts above
$0.3$ (e.g.~Henry 1997) has led to the wide use (e.g~Henry 1997;
Bahcall \& Fan~1998; Eke et al.~1998; Reichart et al.~1999a; Blanchard
et al.~1999; Viana \& Liddle~1999; Donahue \& Voit~1999) of the {\em
EMSS\/} in the estimation of $\Omega_0$; the lack of consensus in the
resulting constraints indicating, at least in part, the difficulty of
using the {\em EMSS\/} data for such a task.

More recently, a number of surveys (Castander et al.~1995; Collins et
al.~1997; Jones et al~1998; Vikhlinin et al.~1998a; Rosati et
al.~1998; Romer et al.~2000) have been created from serendipitous
detections in pointed {\em ROSAT--PSPC\/} observations. These surveys
go much deeper than the {\em EMSS\/}, but over smaller areas: the
largest single survey is the {\em Bright SHARC\/} survey of Romer et
al.~(2000), which covers 179 deg$^2$ to a flux limit of $\simeq2\times
10^{-13}$ erg s$^{-1}$ cm$^{-2}$. The deepest survey, the {\em ROSAT
Deep Cluster Survey, (RDCS)\/} of Rosati et al.~(1998) reaches a flux
limit of $\sim 4 \times 10^{-14}$ erg s$^{-1}$ cm$^{-2}$ over an area
of $\sim$50 deg$^2$.

As we detail below, the {\em XCS\/} will detect much larger numbers of
high--redshift clusters than the existing {\em Einstein\/} and {\em
ROSAT\/} serendipitous surveys. The {\em XCS\/} will benefit not only from
the increased sensitivity of {\em XMM} over {\em Einstein\/} and {\em
ROSAT\/} (see \S\ref{intro}), but also from {\em XMM}'s excellent
spatial and spectral resolution. These advantages would also be shared
by an {\em XMM\/} slew survey of the sort proposed by Jones \& Lumb
(1998).  However, we note that the relatively shallow depth of
an {\em XMM} slew survey ($\sim 2 \times 10^{-13}$ erg s$^{-1}$
cm$^{-2}$) means that it would detect few high redshift clusters, and
thus have little power to discriminate between cosmological models.

\section{A model for the Cluster Population}\label{ps_model}

Our theoretical model for the cluster population uses the extended
Press--Schechter (Press \& Schechter 1974) formalism of Viana
\& Liddle (1999). The validity of this general approach 
has been demonstrated by comparison with $N$-body simulations (Eke,
Cole \& Frenk 1996; Colberg et al.~1998; Tormen 1998). We refer the 
reader to  Viana \& Liddle (1999) for a more detailed description of 
the method.

We consider herein three cosmological models, namely:
\begin{enumerate}
\item the currently--favored spatially-flat, low-density cosmology with
$\Omega_0 = 0.3$ and $\Omega_\Lambda = 0.7$,
\item the Einstein -- de Sitter critical density cosmology, with 
$\Omega_0 = 1$ (and $\Omega_\Lambda = 0$)
\item an open cosmology, with $\Omega_0 = 0.3$ ($\Omega_\Lambda = 0$).
\end{enumerate} 

We note that throughout this paper, we calculate luminosity and
angular diameter distances as follows. For the two $\Omega_\Lambda =
0$ models, we use the standard exact form due to
Mattig (Mattig 1958). For the model with non--zero $\Omega_\Lambda$,
we use the approximate form derived by Pen (1999). This form is
perfectly adequate for our purposes, since it has an error of $\la
1\%$ for flat cosmologies.

We assume that structure formation proceeds through gravitational
instability from a Gaussian distribution of primordial density
perturbations with a scale-invariant power spectrum. The extended
Press--Schechter formalism enables us to compute the number density of
clusters as a function of redshift. The version we use (Viana \&
Liddle 1999) includes a tracking of the merger histories of clusters
in order to account properly for their time of formation when relating
their mass to their temperature. The mass to temperature conversion 
is normalized so as to reproduce the results from the hydrodynamical 
simulations of White et al. 1993b and Bryan \& Norman 1998, with the 
extension to open or flat cosmologies with an arbitrary value for 
$\Omega_{0}$ performed using the expressions given in Viana \& Liddle (1996).
In the following we have only included systems with $T>2$ keV, because the 
Press--Schechter formalism becomes unreliable at low temperatures.

There is a weak dependence of Press--Schechter results on the current
shape of the linear power spectrum of density fluctuations, so, for
definiteness, we have used a Cold Dark Matter (CDM) power spectrum
with shape parameter $\Gamma$ (Efstathiou, Bond \& White 1992) equal
to $0.23$, as suggested by some analyses of galaxy clustering (Peacock
\& Dodds 1994; Viana \& Liddle 1996; but see Mann, Peacock \& Heavens
1998). The normalization of the power spectrum is that of Viana \&
Liddle (1999), ensuring that the present-day abundance of
high-temperature clusters is recovered. Similar cluster-based normalizations 
were also obtained by Eke, Cole \& Frenk 1996, Pen 1998, Borgani et al. 1999,
Blanchard et al. 1999 and Henry 2000. Further, these models give a good
fit to the {\em COBE\/} 4--year data (e.g.~Tegmark 1996).

\begin{figure*}
\centering
\leavevmode\epsfysize=7.cm \epsfbox{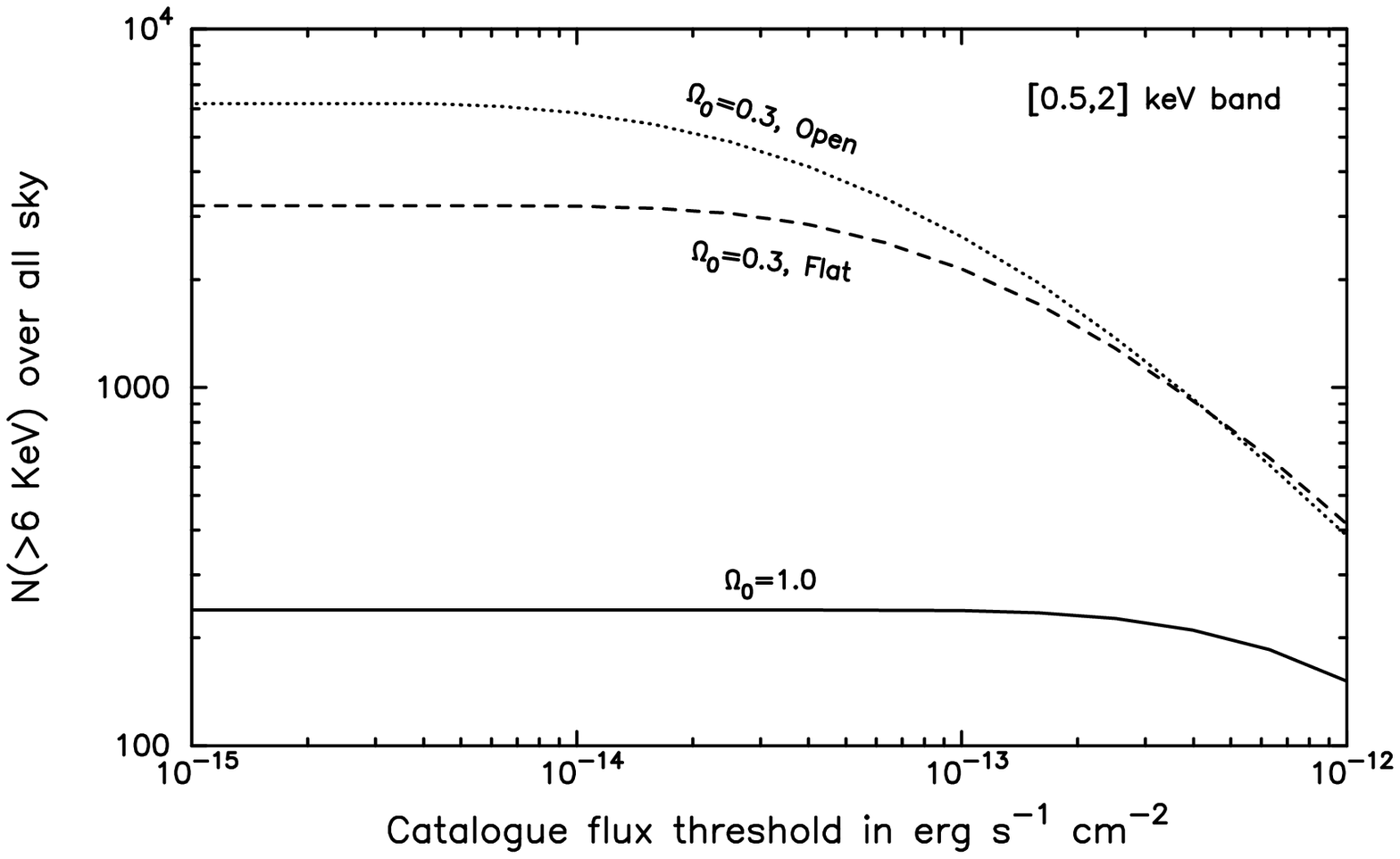}\\
\leavevmode\epsfysize=7.cm \epsfbox{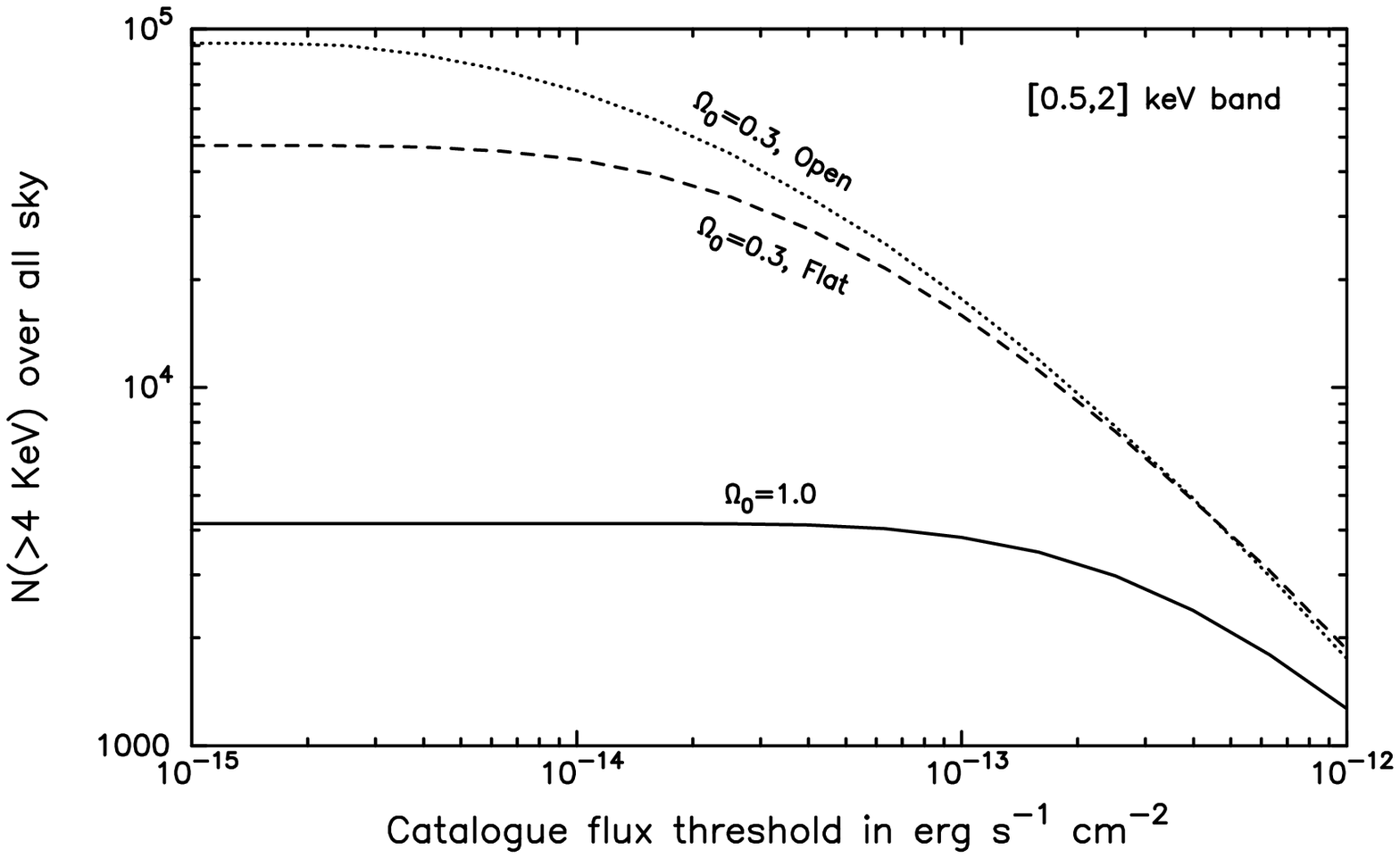}\\
\leavevmode\epsfysize=7.cm \epsfbox{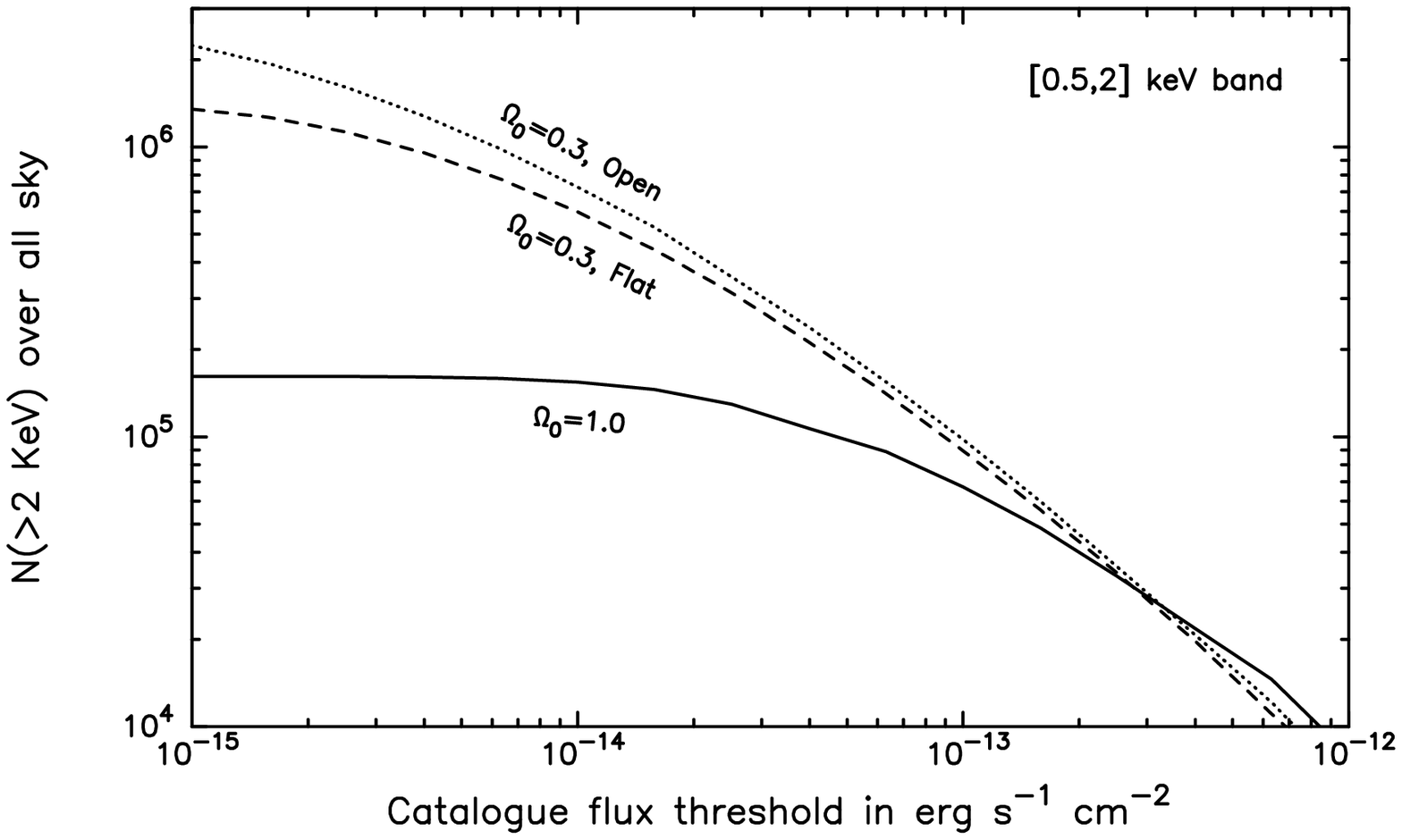}\\
\caption{\label{fluxallz} The expected number of galaxy clusters
across the whole sky with X-ray temperatures in excess of 6 keV (upper
panel), 4 keV (middle panel) and 2 keV (lower panel), as a function of
the catalogue flux threshold in the 0.5-2.0 keV band. The
$\Omega_{0}=1$ case is the solid line, while for $\Omega_{0}=0.3$ the
flat case is shown as dashed and the open case as dotted.}
\end{figure*}

In Fig.~\ref{fluxallz} we plot the cumulative number, $N$, of clusters
with temperatures greater than 2, 4, and 6 keV in the whole sky as
function of flux cut, $f$, for the three cosmologies. To derive the
$N(f)$ functions from the Press--Schechter results we have to assume a
conversion from cluster temperature to luminosity. For this we use the
empirical cluster luminosity--temperature relation derived by Allen \&
Fabian (1998, AF98);
\begin{equation}
T = 1.66 L_{\rm X}^{0.429} 
\label{l_t_af98}
\end{equation} 
where the temperature, $T$, is in keV and the bolometric luminosity,
$L_{\rm X}$, in is units of $10^{44} \, {\rm erg \, s}^{-1}$.
Observations to date present no evidence for significant evolution of
the $L_{\rm X}-T$ relation out to $z \sim 0.4$ (AF98; Mushotzky \&
Scharf~1997; Reichart, Castander \& Nichol 1999b), 
but nothing is known beyond that.  For the
purposes of our calculations, we assume that equation~(\ref{l_t_af98})
holds at all redshifts, but stress (as discussed further in
\S\ref{discussion}) that one of the principal scientific results of
the {\em XCS\/} will be a greatly improved understanding of the
$L_{\rm X}-T$ relation and its evolution with redshift.

In Press--Schechter theory, the {\em relative} abundance of galaxy
clusters of a given mass at two given redshifts depends only on the
growth rate of perturbations, which in turn depends only on $\Omega_0$
and $\Omega_\Lambda$. Fig.~\ref{fluxallz} shows that $N(f)$ varies
between the three cosmologies more dramatically as temperature
increases (note the different scales in the three panels). Moreover,
below $T\simeq4$ keV, it is possible that the mass-temperature
relation has been significantly influenced by heat injection into the
intergalactic medium. For these reasons, we will largely focus our
discussion on clusters with X-ray temperatures in excess of 4 keV (or
luminosities $\gtrsim 2.6 \times10^{44}$ erg s$^{-1}$, based on the
AF98 $L_{\rm X}-T$ relation). Although we note that, in practice, 
the optimum (i.e. the one that minimizes the errors on cosmological 
parameter estimates) temperature limit for the {\em XCS} will probably 
not be exactly 4 keV.

\begin{figure*}
\centering
\leavevmode\epsfysize=7.cm \epsfbox{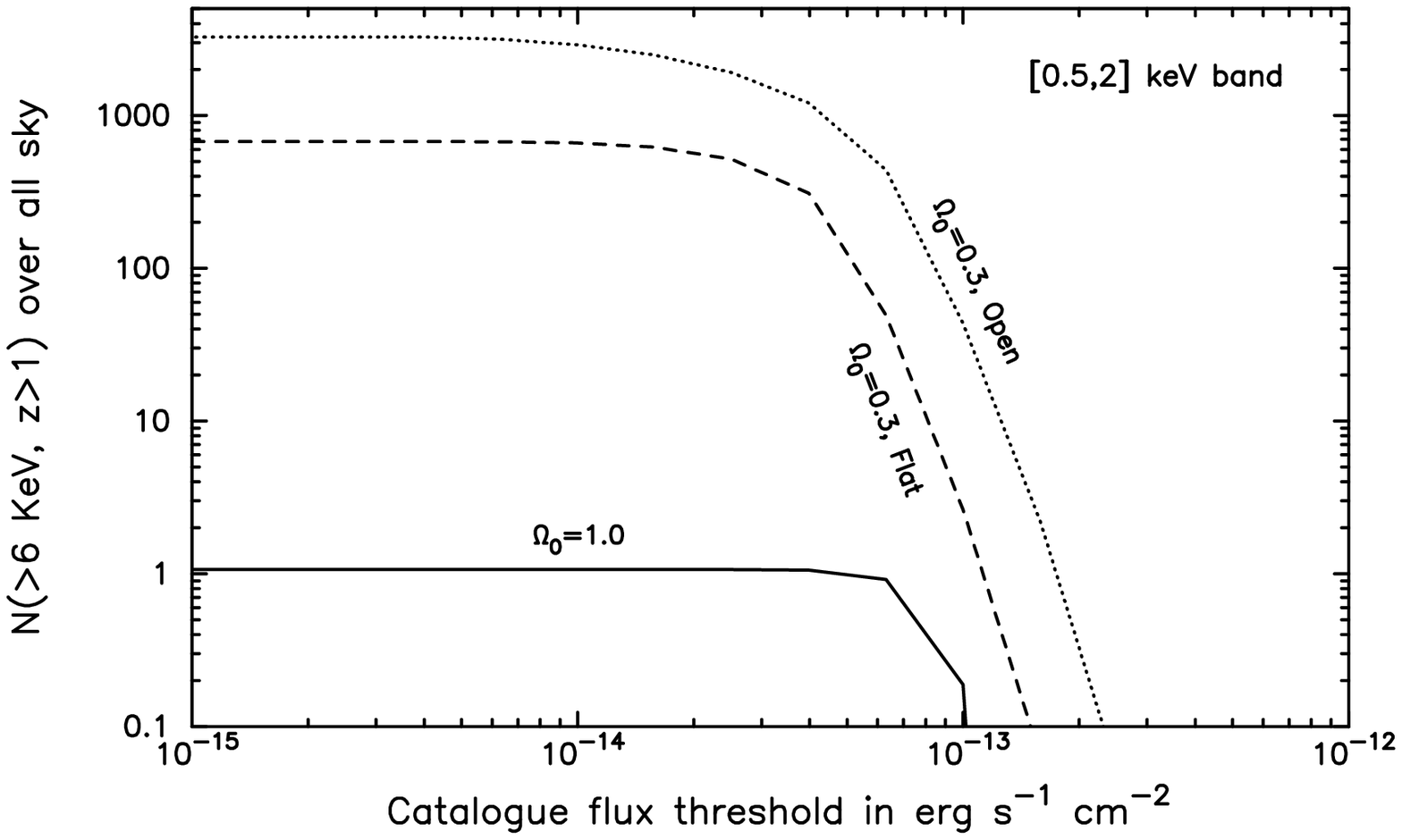}\\
\leavevmode\epsfysize=7.cm \epsfbox{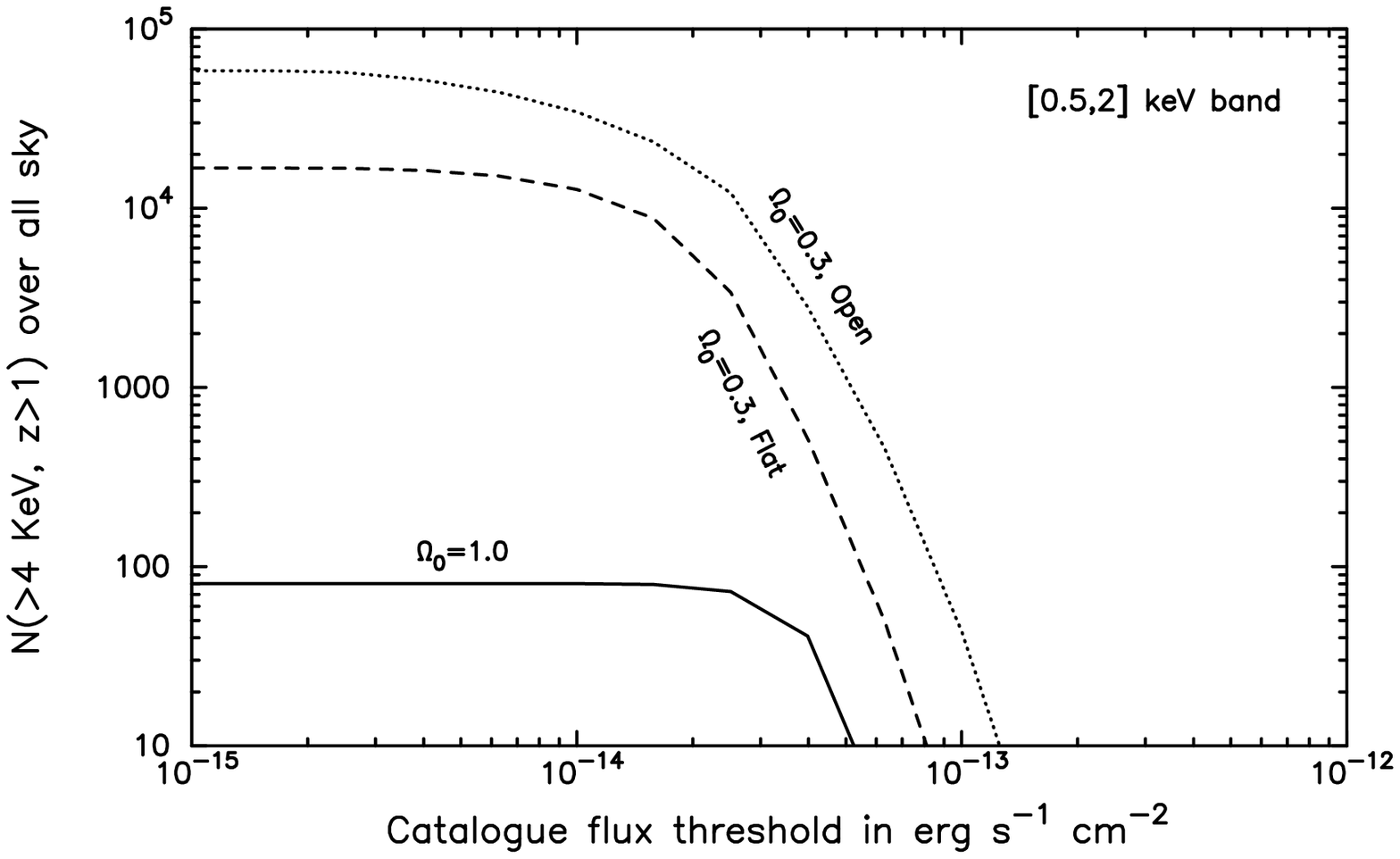}\\
\leavevmode\epsfysize=7.cm \epsfbox{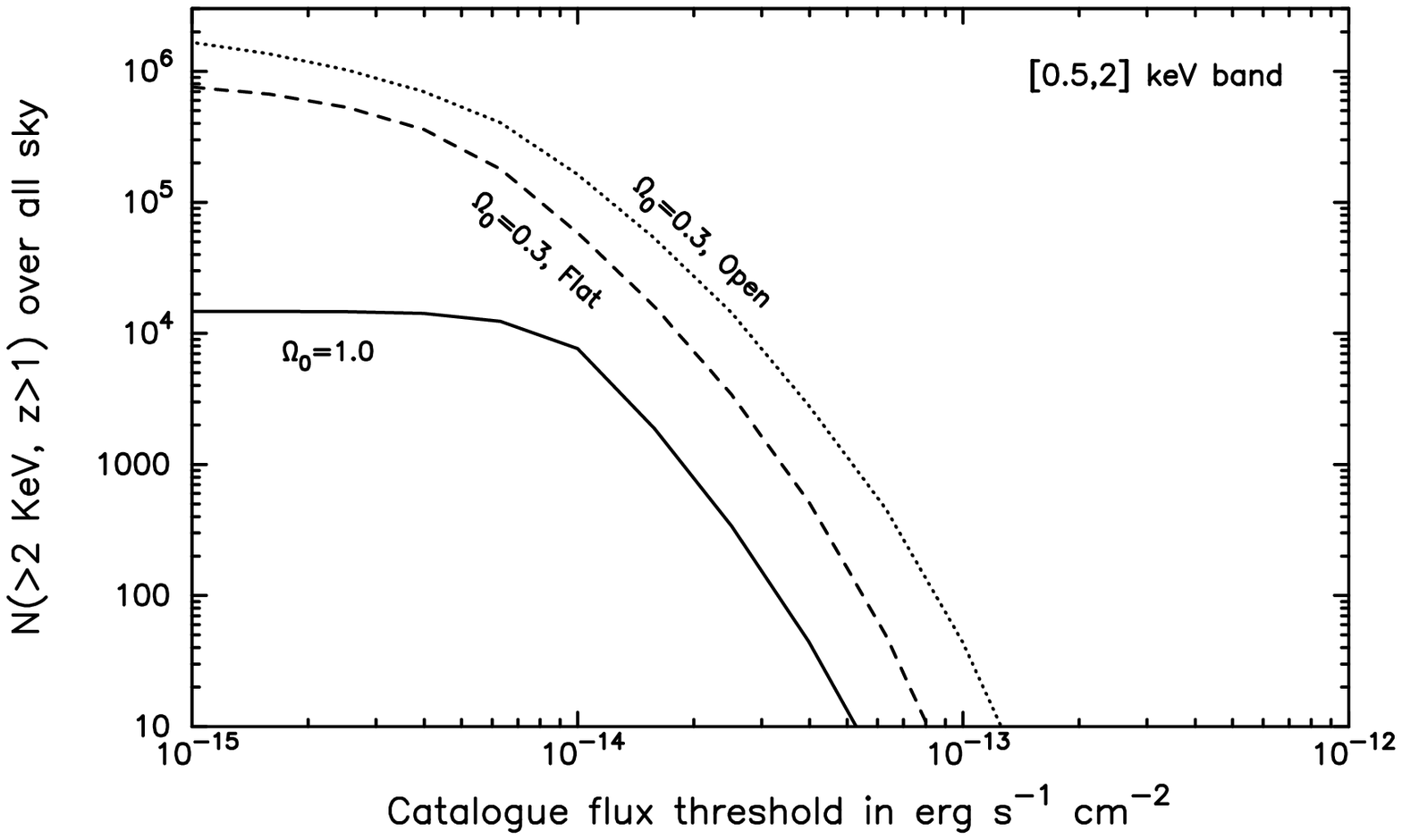}\\
\caption{\label{fluxgt1} As Fig.~\ref{fluxallz}, but showing only
clusters with $z>1$.}
\end{figure*}

\begin{figure*}
\centering
\leavevmode\epsfysize=7.cm \epsfbox{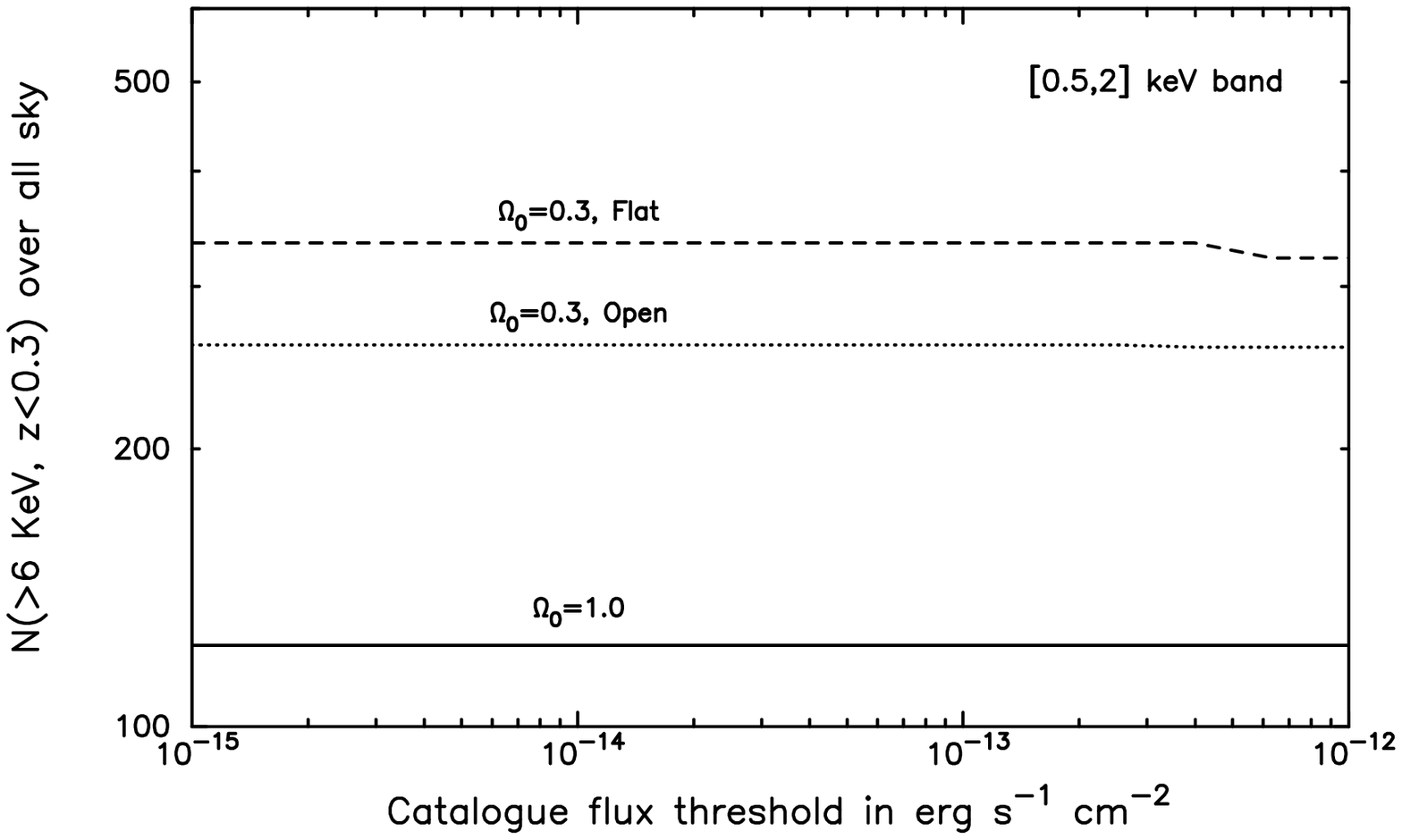}\\
\leavevmode\epsfysize=7.cm \epsfbox{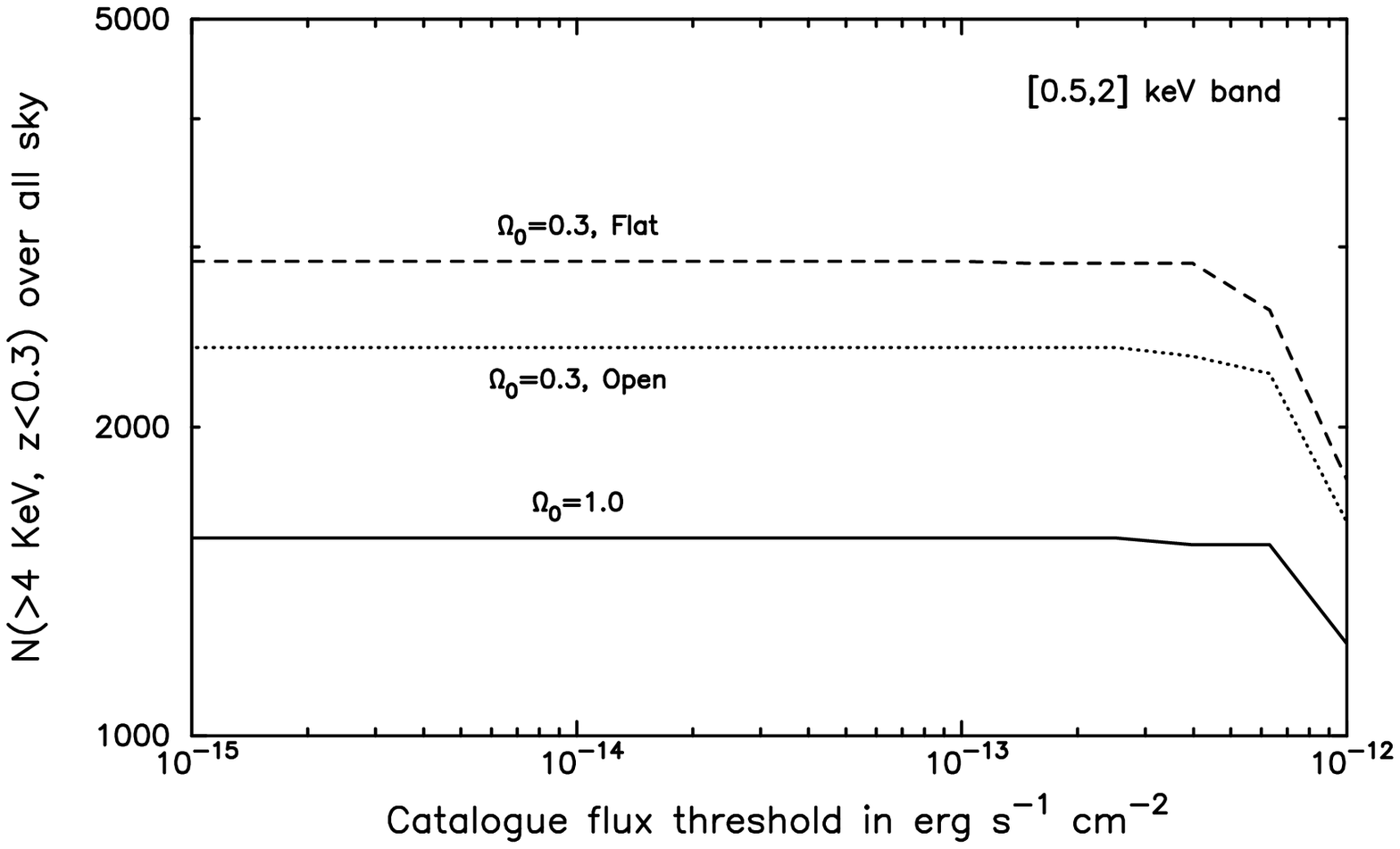}\\
\leavevmode\epsfysize=7.cm \epsfbox{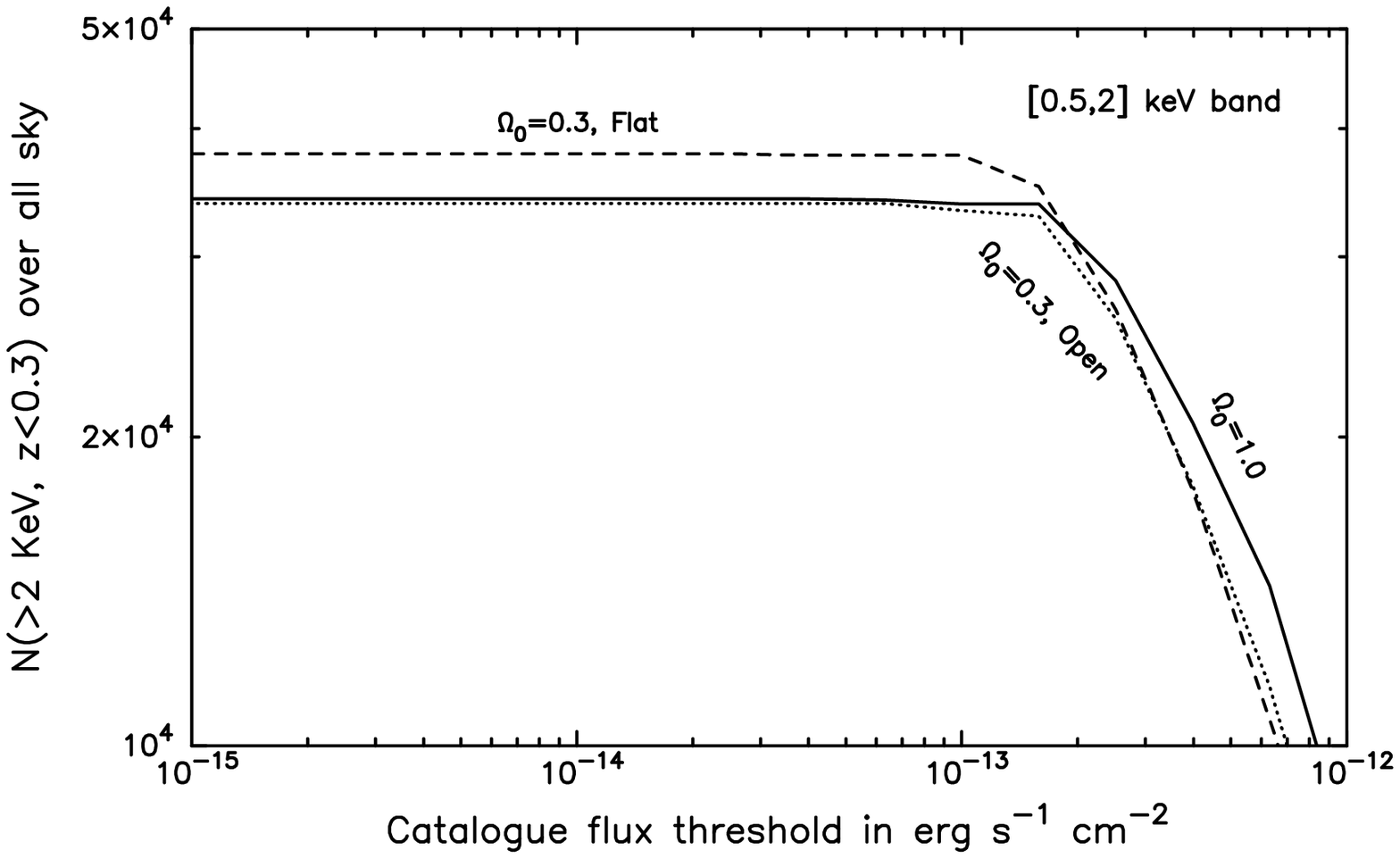}\\
\caption{\label{fluxlt0.3} As Fig.~\ref{fluxallz}, but showing only
clusters with $z<0.3$.}
\end{figure*}

From Fig.~\ref{fluxallz} it is clear that distinguishing between high
and low values of $\Omega_0$ is relatively straight forward, but that
discriminating between open and flat models with the same value of
$\Omega_0=0.3$ is much harder. To do so, one needs to have access to
clusters at sufficiently high redshift, as illustrated by
Fig.~\ref{fluxgt1} which shows the cumulative flux distribution of
clusters at $z>1$: the predicted numbers of $T>6$ keV clusters at $z>1$
in the two $\Omega_0=0.3$ models differ by as much as a factor of four
at faint flux limits, compared to less than than a factor of two for
$z>0$.  In Fig.~\ref{fluxlt0.3} we plot the analogous curves for
clusters at $z<0.3$, to demonstrate that cluster catalogues limited to
low redshift (such as those produced by the various {\em
RASS\/}--based surveys or that to be produced by the Sloan Digital Sky
Survey of Gunn et al.~1998) are very poor at constraining
cosmological parameters: it is necessary to reach $z\ga0.5$ to get a 
sufficiently long lever arm in cosmological time for the sensitivity of
the growth rate of density perturbations to cosmology to become apparent.

The curves plotted in Fig.~\ref{fluxlt0.3} are flat for fluxes fainter
than $\sim10^{-13}$ erg s$^{-1}$ cm$^{-2}$, indicating that all
$z<0.3$ clusters with temperatures above $T=2$ keV can be detected
above that flux limit. This is no surprise, of course: by assumption
(through equation~\ref{l_t_af98}) clusters above a certain temperature
also exceed a certain luminosity, so there clearly must be a flux
level at which they are all visible if a redshift limit is imposed.
What is more interesting is that this asymptoting behavior is also
seen in the curves in Fig.~\ref{fluxallz}, for which there is no
redshift limit. The levelling off of the curves in Fig.~\ref{fluxallz}
results from the fact that clusters do not exist at arbitrarily high
redshifts in a hierarchical universe: it takes a certain amount of
time to accumulate the matter making up a cluster of a given mass
(and, hence, temperature and luminosity). So, if one can reach a
sufficiently faint flux limit, one can look along one's past
light--cone beyond the epoch when the first cluster of a particular
mass was formed. For the high temperature ($T>4$ keV) clusters
important for cosmological parameter estimation, this asymptote is
being reached at a depth ($\sim 10^{-14}$ erg s$^{-1}$ cm$^{-2}$)
which is comparable to that to be reached by the {\em XCS\/}
(\S\ref{results:catalogues}). This implies that there would be no
point in ever performing a deeper survey than the {\em XCS\/} if the
sole purpose of that survey was to detect $T>4$ keV clusters; only a
survey with a wider sky coverage would yield more detections (and
hence tighter constraints on $\Omega_0$ and
$\Omega_\Lambda$). Although we note that, by going deeper, one can
obtain useful information about individual clusters, such as their
spatial morphology and their temperature profile.

\section{Assumptions and Simplifications made when Simulating the 
Sensitivity Limits of the {\em XCS\/}}\label{assumptions}

In \S\ref{ps_model} we predicted the number of clusters on the sky,
$N$, as a function of X--ray flux, $f$.  If clusters were the only
sources in the X--ray sky, then $N(f)$ would describe the cluster
catalogue resulting from the idealized situation of an all-sky survey
performed by an instrument with no internal background, a vanishingly
narrow point spread function (so that confusion noise is zero) and a
uniform flux limit. In reality, the {\em XCS} will have a non-uniform
flux limit, will only cover a fraction of the sky and will have to
detect clusters against a significant X-ray background.  Moreover,
since the X--ray sky is dominated by point sources (e.g. AGN), a
crucial step in its construction will be the differentiation of
point-like from extended sources.

So, in order to predict how many, and what type of, clusters will
be included in the {\em XCS\/} catalog, we need to make various 
assumptions about the sensitivity and spectral response of the 
instrument; about the surface brightness profiles and spectral
properties of the clusters we expect to observe; and about the 
properties of the X-ray background. We detail these, and other,
assumptions below.

\subsection{EPIC--pn only}\label{pn_only}

We have simplified our calculations by concentrating only on the
EPIC--pn camera.  This is because the EPIC--MOS cameras receive only
50\% of the flux from their respective telescope modules (the other
50\% in each goes to an RGS), and because the MOS CCDs are
intrinsically less sensitive than the pn CCDs. An additional advantage
is that the simulation of the catalogue selection function will also
be simplified if data from only one camera are used.  However, once
the clusters have been detected, the EPIC--MOS data can be used to
help parameterize the cluster morphology and spectrum.  By prudent use
of the EPIC--MOS data, the percentage of clusters with accompanying
temperatures should increase over that suggested by
Table~\ref{cluster-numbers}.

\subsection{Minimum Detection Threshold of 8$\sigma$}\label{8sigma}

We adopt a minimum detection threshold of $8 \sigma$. This is because
the {\em XCS} will have to rely on source extent to differentiate
clusters from point-like X-ray sources, such as stars and AGN, and it
has been shown (e.g. Greg Wirth, private communication) that extent
measures can only be derived with confidence for sources detected at
$\gtrsim8 \sigma$.  The {\em Bright SHARC} survey (Romer et al.~2000)
also used a minimum detection threshold of $8 \sigma$ for this reason.

\subsection{Detection Significance Computed using Inner 50\% of Flux} 
\label{inner50}

When calculating the detection significance, we only consider the
inner 50\% of the total cluster flux. (We define the radius of the
region enclosing this flux as $r_{50}$.) This is because automated
source detection algorithms tend to underestimate the count rates of
extended sources. For example, the wavelet transform method adopted by
the {\em Bright SHARC} survey, to analyse {\em ROSAT} data,
underestimates the count rate of $z>0.15$ clusters by a factor of 2.1
(Romer et al.~2000). This is a conservative assumption: we would hope
that more efficient cluster selection algorithms will be developed, to
make full use of the higher quality {\em XMM\/} data.

\subsection{Clusters Follow Spherically--Symmetric Isothermal $\beta$=2/3 
Model}\label{beta_model}

To estimate $r_{50}$ values, we assume that all clusters can be
modeled as spherically-symmetric systems that follow an isothermal
$\beta$-profile
\begin{equation}
I=\frac{I_0}{[1+(r/r_c)^2]^{3\beta -1/2}} \,\,,
\label{beta-eq}
\end{equation}
where $I$ is the surface brightness at radius $r$, $r_c$ is the core
radius and $-3\beta$ is the asymptotic radial fall-off of the ICM
(intracluster medium) density distribution. We adopt $\beta=2/3$
throughout, since this is a typical value for rich clusters (Jones \&
Forman 1984, Mohr et al. 1999), although we note that its value 
for any given cluster can vary in the range $0.4\lesssim\beta\lesssim0.9$. 
For a cluster described by equation~(\ref{beta-eq}), $\beta=2/3$ gives
$r_{50}=\sqrt{3}r_c$. We adopt this model for the cluster surface
brightness, as it has been shown (e.g. Mohr et al.~1999) to describes 
the {\em azimuthally-averaged\/} cluster emission in the ROSAT band-pass
(0.5-2.0 keV) very well. However, as we discuss in \S~\ref{cluster_limits},
the use of such a simplistic model is one of the major limitations of
our calculations.

\subsection{Non--Evolving Core Radius -- Luminosity Relation}\label{rc_l44}

We further assume that the core radius follows the relation
\begin{equation}
r_c= \frac{250}{h_{50}}\left(\frac{L_{44}}{5}\right)^{0.2} {\rm kpc} \,\,,
\label{lum-eq}
\end{equation}
where $L_{44}$ is the rest-frame luminosity in the 0.5-2.0 keV band in
units of $10^{44}$ erg s$^{-1}$. This relation was proposed by Jones
et al.~(1998) and has been shown to agree with measured values of
$r_c$ for clusters with luminosities in the range $10^{43}$ erg
s$^{-1}$ to $10^{45}$ erg s$^{-1}$. We assume that the core radius
does not evolve (as shown by Vikhlinin et al.~1998b). A better
understanding of the luminosity -- core radius relation (in
particular, whether it evolves with redshift) should result from
forthcoming observations of known clusters with {\em Chandra\/} and
{\em XMM\/}.

\subsection{Cluster Count Rates}\label{cluster_count_rate} 

To determine how the {\em XMM} count rate varies with cluster parameters, we
used the {\tt fakeit} and {\tt show rates} commands in {\sc xspec} and
assumed that the cluster X-ray emission can be described by absorbed
Raymond--Smith spectra. We calculated unabsorbed fluxes and on-axis
{\em XMM} count rates for spectra with 12 different temperatures
($1<T<12$ keV, in 1 keV increments) and 40 different redshifts
($0.05<z<2.0$, in $\Delta z=0.05$ increments); 480 spectra in all.
Throughout we kept the metallicity fixed at $Z=0.3Z_{\odot}$ (see
\S\ref{metallicity}), the normalization fixed at 1 and the Galactic
{\sc hi\/} column density fixed at $n_H=4\times10^{20}$ cm$^{-2}$ (see
\S\ref{column-density}). The {\tt ignore\/} command was used to limit
the count rate calculation to the 0.5-10 keV band. (The full energy
range over which EPIC--pn is sensitive is 0.1-11 keV.) 

These 480 calculations provided us with the count rate to flux
conversion factors that were used to define the survey sensitivity
limits in \S\ref{results}. To illustrate how these conversion factors
vary with $T$ and $z$, we provide some examples: A Raymond--Smith
spectrum with an unabsorbed flux of $1\times10^{-13}$ erg s$^{-1}$
cm$^{-2}$ will yield 0.078 EPIC--pn counts s$^{-1}$ (0.5-10 keV) when
$T=1$ keV and $z=0.1$.  A spectrum with the same flux will yield 0.071
counts s$^{-1}$ when $T=1$ keV and $z=1$, 0.100 counts s$^{-1}$ when
$T=10$ keV and $z=0.1$, and 0.093 counts s$^{-1}$ when $T=10$ keV and
$z=1$. 

We note that the AF98 $L_{\rm X}-T$ relation used in section \S\ref{ps_model}
was constructed using a slightly different plasma model to that used
here; {\tt mekal} (Kaastra \& Mewe 1993) in {\sc xspec} rather than
{\tt raymond}.  However, this does not present a problem for this
study since we limit our discussion to $T>2$ keV clusters (the
predictions of {\tt mekal} and {\tt raymond} are very similar above
$T\gtrsim1$ keV).

\subsection{Cosmic Background Count Rate}\label{cosmic_background}

We calculated the cosmic background using a model that includes two
thermal Galactic components (modeled with absorbed Raymond--Smith
spectra) and a power-law extragalactic component. The first thermal
component had a temperature of 0.0258 keV, a metallicity of
$Z_{\odot}$, a redshift of $z=0$, a normalization of $2.5 \times
10^{-6}$ and a hydrogen column density of $n_H = 1 \times 10^{17}$
cm$^{-2}$ (Labov \& Bowyer~1991). The second thermal component had a
temperature of 0.0947 keV, a metallicity of $Z_{\odot}$, a redshift of
$z=0$, a normalization of $3.0 \times 10^{-6}$ and a hydrogen column
density of $n_H = 6\times10^{19}$ cm$^{-2}$ (Rocchia et al.~1984). The
power--law component had an index of $\alpha=1.4$, a normalization of
$9.32 \times 10^{-7}$ (Chen et al.~1997) and a hydrogen column density
of $n_H=4\times10^{20}$ cm$^{-2}$. The adopted cosmic background model
yields a count rate of $2.6\times10^{-3}$ s$^{-1}$ arcmin$^{-2}$ in
the 0.5-10 keV band. We note that, when calculating the
signal--to--noise of cluster detections, we adjust the cosmic
background count rate by the appropriate vignetting factor (see
\S\ref{vignetting}). The true external background (made up of solar,
Galactic and extragalactic components) is known to vary considerably
across the sky, but most of this variation is confined to low energies
($<1$ keV, Snowden et al.~1997), and so this should not have a
significant effect on the average signal--to--noise values for cluster
detections we calculate in the 0.5-10 keV band.

\subsection{Particle Background Count Rate}\label{particle_background}

We calculated the particle background using the expected internal
background rates quoted in the {\em XMM Users' Handbook\footnote {{\tt
astro.estec.esa.nl/XMM/user/uhb\_top.html}}\/}: $3.0\times10^{-4}$
counts cm$^{-2}$ s$^{-1}$ keV$^{-1}$ for the EPIC--pn detector. The
spectrum of the internal background is expected to be flat, so the
integrated count rate in the 0.5-10 keV band is
$9.5\times3.0\times10^{-4}$ counts cm$^{-2}$ s$^{-1}$. We then
converted from cm$^{-2}$ to arcmin$^{-2}$ to obtain a rate of
$1.4\times10^{-4}$ counts s$^{-1}$ arcmin$^{-2}$ ($4.1''$ corresponds
to 150 $\mu$m at the detector).

\subsection{Vignetting Correction}\label{vignetting}

The count rate to flux conversion factors calculated using {\sc xspec} 
(\S\ref{cluster_count_rate}) refer to the on-axis response of the EPIC--pn.
In order to account for how these conversion factors vary with
off-axis angle, we had to calculate vignetting corrections. We did this 
as follows; using the {\sc quicksim}\footnote{Available from {\tt
legacy.gsfc.nasa.gov}} package written by Steve
Snowden, we created fake EPIC--pn images of a
point source, with a Raymond--Smith spectrum, in the absence of
particle and cosmic backgrounds. By placing the source at various
places in the field of view, we were able to measure how the count
rate varied as a function of off-axis angle.  The vignetting factor
changes smoothly across the field of view, so we decided to break up
the field of view into five $3'$ wide annuli ($\bar\theta=1'.5,4'.5,7'.5,10'.5,
13'.5$ respectively).  
 For each annulus we calculated the mean vignetting factor for a point
 source with a $T=4$ keV spectrum; this was found to be 0.987, 0.892,
 0.734, 0.578 and 0.520 respectively.  We used a single temperature
 for this calculation because we found the vignetting factor to be
 essentially independent of temperature; the on-axis sensitivity is
 2.09 times that of the sensitivity at $\theta=12'$ for a $T=1$ keV
 spectrum, compared to 2.14 for a $T=8$ keV spectrum. We also made a
 mega-second {\sc quicksim} simulation of the cosmic background, in
 the absence of sources and a particle background, to confirm that
 these average vignetting factors also apply to the cosmic background.
 
\subsection{Constant ICM Metallicity: $Z=0.3Z_\odot$}\label{metallicity}

The X--ray emission from an astrophysical plasma is a function of its
metallicity.  For example, we calculate that for a Raymond--Smith
spectrum with an unabsorbed flux of $1\times10^{-13}$ erg s$^{-1}$
cm$^{-2}$, the count rate varies from 0.077 s$^{-1}$ to 0.078 s$^{-1}$
to 0.079 s$^{-1}$ for $Z=0.1Z_{\odot}$, $0.3Z_{\odot}$ and $Z_{\odot}$
respectively ($T=1$ keV, $z=0.1$, $n_H=4\times10^{20}$ cm$^{-2}$).
Because of this weak dependence of count rate to metallicity, we adopt
a constant value of $Z=0.3Z_{\odot}$, as this is typical of rich
clusters: Fukazawa et al.~(1998) found that the ensemble-averaged iron
abundance was $0.3\pm0.02$ based on {\em ASCA} observations of 40
nearby clusters of galaxies. Further, we assume that metallicity does
not evolve; up to $z\sim1$ there is observational support for this
from Tsuru et al.~(1997) and Schindler~(1999), and theoretical support
from calculations by Martinelli et al.~(2000).

\subsection{Constant {\sc hi} Column Density: $n_{\rm H}=4\times10^{20}$ 
cm$^{-2}$}\label{column-density}

Neutral hydrogen gas along the line of sight towards a cluster,
particularly within our own galaxy, absorbs a large fraction of the
emitted X-rays at low ($\la$ 0.5 keV) energies. Since we do not know
what the actual distribution of hydrogen column densities will be in
the {\em XCS}, we have adopted a single value, $n_H=4\times10^{20}$
cm$^{-2}$, which is typical for high Galactic latitudes.

The effect of column density on count rates is not large.
For Raymond--Smith spectra with unabsorbed fluxes of
$1\times10^{-13}$ erg s$^{-1}$ cm$^{-2}$, the count rate varies from
0.086 s$^{-1}$ to 0.078 s$^{-1}$ to 0.064 s$^{-1}$ for $n_H$ equal to
1, 4 and 10 $\times10^{20}$ cm$^{-2}$ respectively ($Z=0.3Z_{\odot}$,
$T=1$ keV, $z=0.1$).  Our adoption of 4 $\times10^{20}$ cm$^{-2}$ is
on the conservative side; many of the regions explored by {\em XMM\/}
will have lower $n_H$ values. For example, of the 37 clusters in the
Bright {\em SHARC} survey (Romer et al.~2000), all but 9 were detected
in regions with $n_H < 4 \times10^{20}$ cm$^{-2}$.  This means that
the number of clusters eventually detected by the {\em XCS\/} could
well be higher than suggested by Table~\ref{cluster-numbers}.

\subsection{EPIC Thin Filter used for all Observations}\label{blocking_filter}

The EPIC cameras are sensitive not only to X--rays, but also to
optical photons. Optical blocking filters (thin, medium or thick) are
used to minimize the number of photons entering the detector. For our
calculations we use the response functions corresponding to the thin
filter only.

The choice of optical filter has an even smaller effect on count rates
than column density. For a Raymond--Smith spectrum with an unabsorbed
flux of $1\times10^{-13}$ erg s$^{-1}$ cm$^{-2}$, the count rates vary
from 0.078 s$^{-1}$ to 0.076 s$^{-1}$ to 0.06 s$^{-1}$ when the thin,
medium and thick filters are respectively in place ($Z=0.3Z_{\odot}$,
$T=1$ keV, $z=0.1$, $n_H=4\times10^{20}$ cm$^{-2}$). To calculate the
count rate through the thin, medium and thick filters we used the
files epn\_thin\_arf.fits, pn\_med\_arf.fits and
epn\_thick\_arf.fits\footnote{All available from {\tt
astro.estec.esa.nl} in the directory tree {\tt
/pub/XMM/EPIC/March99/RESPONSES}} respectively.  It is unlikely that
any {\em XMM\/} pointings that require the thick filter (i.e.~those
with bright stars in their field of view) will be suitable for
serendipitous cluster detection and so we can safely discount the
effects of filter choice on the cluster numbers presented in
Table~\ref{cluster-numbers}.

\subsection{Bolometric and K-corrections}
\label{bkcorr}

We calculated K-corrections and bolometric corrections using {\sc
xspec}.  The K-correction was defined as the ratio of the unabsorbed
flux in the observed energy band to the unabsorbed flux in the
redshifted energy band
\begin{equation}
K_{lo-hi}=\frac{\int^{hi}_{lo}f_{\nu}d\nu}{\int^{hi(1+z)}_{lo(1+z)}
f_{\nu}d\nu} \,,
\label{k-corr}
\end{equation}
where $lo$ and $hi$ are the limits of the observed energy band,
e.g.~0.5 and 10.0 keV.  When calculating K-corrections, the redshift
of each Raymond--Smith spectrum was set to $z=0$. Quadratic fits to
the K-corrections, as a function of $(1+z)$ were derived for each of
the input temperatures (1 to 12 keV);
\begin{equation}
K_{lo-hi}=c+b(1+z)+a(1+z)^2\,\,, 
\label{k-quad}
\end{equation}
where $a,b,c$ are the coefficients of the fits, see
Table~\ref{k-corr-tab}.

The bolometric correction was defined as the ratio of the unabsorbed
flux in a pseudo-bolometric band of 0.01-50 keV to the unabsorbed flux
in the observed energy band, i.e.;
\begin{equation}
B_{lo-hi}=\frac{\int^{50}_{0.01}f_{\nu}d\nu}{\int^{hi}_{lo}
f_{\nu}d\nu}\,\,.
\label{b-corr}
\end{equation}
Setting the redshift of the Raymond--Smith spectrum to $z=0$, $B$
values were calculated for each of the 12 input temperatures. The
bolometric corrections are listed in Table~\ref{b-corr-tab} for the
0.5-2.0 keV and 0.5-10 keV energy bands.

To illustrate how the bolometric and K-corrections were applied, we
provide an example. Consider a cluster with temperature $T=4$ keV, a
redshift of $z=1$, and an unabsorbed flux in the 0.5-2.0 keV
(observed) band of $1\times 10^{-13} \, {\rm erg\, s}^{-1} \, {\rm
cm}^{-2}$.  The K-corrected flux in the 0.5-2.0 keV (rest frame) band
is $0.836 \times 10^{-13} \, {\rm erg\, s}^{-1} \, {\rm cm}^{-2}$
(from equation~\ref{k-quad} and Table~\ref{k-corr-tab}). The
bolometric flux for this cluster is then $3.04$ times this, from
Table~\ref{b-corr-tab}.
       
\subsection{Exposure Time Distribution and Area of the Survey}\label{extdist}

In \S\ref{results:catalogues} we combine our sensitivity limit
calculations (\S\ref{temp-est} \& \S\ref{detectcalc}) with our model
cluster population (\S\ref{ps_model}) in order to predict the
properties of the {\em XCS}.  To do so requires us to assume both an
areal coverage and an exposure time distribution for the survey.  We
do not know what the exposure time distribution will be for the
thousands of pointings that will eventually comprise the {\em XMM}
archive. So, for the purposes of this paper, we assume that the
exposure times will be distributed in the same way as they are for 760
pointings in the {\em XMM\/} Guaranteed Time Observations (GTO,
Table~\ref{exp-dist}). The 760 GTO pointings have exposure times that
range from 5 ks to 95 ks, with an average of 22.3 ks. For comparison,
we also list in Table~\ref{exp-dist} the distribution of the exposure
times in the {\em XMM\/} A01 program.

For the areal coverage, we use a total value of $800$ deg$^{2}$ (as
justified below). However, we note that our treatment of vignetting
effects (\S\ref{vignetting}) forces us to break this total area up
into five bins when creating mock cluster catalogues. These bins
correspond to the five adopted off-axis annuli, which cover 4.3\%,
13.1\%, 21.7\%, 30.4\% and 30.2\% of the total area respectively.

The EPIC field of view covers a $30'$ diameter circle and the CCD
arrangement of the pn camera provides an active area of 649 square
arcminutes. If EPIC operates for the full ten years of the {\em XMM}
design lifetime, and {\em XMM} makes an average of three pointings per
day, then the total area imaged by EPIC will be $\simeq2000$
deg$^2$. (Although the average exposure time of the 760 GTO pointings
is 22.3 kilo-seconds, or 3.9 pointings per day, overheads, such as the
$\sim 5$ ks telescope settling time, mean that three pointings per day
is a more realistic estimate.)  Unfortunately, not all of the
$\simeq2000$ deg$^2$ will be available for building serendipitous
cluster catalogues.  Experience from {\em ROSAT\/} suggests that only
$\sim40$\% of pointings are likely to be suitable, the rest being
either at low Galactic latitude, overlapping previously studied
fields, or have pointing targets extending over most of the field of
view. Therefore, we estimate that the {\em XCS} will cover $\simeq
800$ deg$^2$.

\section{Simulation Results}\label{results}

The mechanisms outlined in \S\ref{assumptions} allow us
to simulate the sensitivity of the {\em XCS} in terms of both 
cluster detection (\S\ref{detectcalc}) and temperature estimation 
(\S\ref{temp-est}).
The results of these simulations, when combined with the Press--Schechter
predictions described in \S\ref{ps_model}, allow us to predict
the properties of the {\em XCS} (\S\ref{results:catalogues}).

\subsection{Calculation of Sensitivity Limits for Detection}
\label{detectcalc}

As stated in \S\ref{8sigma}, our criterion for source detection is
that it should be made with count statistics significant at the
8$\sigma$ level at least, so that it is possible to determine whether
the source is extended or not. For the {\em XCS} predictions, we have
calculated the bolometric luminosity that would yield an 8$\sigma$
detection ($L_{8\sigma}$) for each of 144,000 different parameter
combinations. These 144,000 combinations comprise of 3 cosmologies
(\S\ref{ps_model}), 12 temperatures ($1<T<12$ keV, in 1 keV
increments), 40 redshifts ($0.05<z<2$, in $\Delta z=0.05$ increments),
5 off-axis angles ($\bar\theta=1'.5,4.'5, 7'.5,10'.5,13'.5$,
\S\ref{vignetting}) and 20 exposure times ($5<t<100$ ks, in 5 ks
increments).

We determine the 144,000 $L_{8\sigma}$ values iteratively as follows.
For a particular $\Omega_0$, $\Omega_\Lambda$, $T$, $z$, $\theta$ and
$t$ combination, we start by calculating the half--flux radius,
$r_{50}$ (\S\ref{inner50}), in arcminutes for a given input $r_c$
value ($r_{50}=\sqrt{3}r_c$, \S\ref{beta_model}).  Next, we calculate
the total number of background counts, $N$,
(\S\ref{cosmic_background}, \S\ref{particle_background}) that would
fall in a circle of radius $r_{50}$ in the exposure time $t$ (where
$N$ takes into account the effects of vignetting on the cosmic
background at off-axis angle $\theta$).  Once $N$ is known, we can 
calculate the number of cluster counts, $S$, that would
need to fall inside $r_{50}$ to yield $S/N>8$.  By multiplying $S$ by
$2/t$, we obtain the corresponding total count rate (where the factor of 2 
accounts for those photons lying outside the $r_{50}$ radius and the
factor of 1/t converts from total counts to a count rate).  This count rate
can then be converted into a flux by dividing by the appropriate
vignetting factor (\S\ref{vignetting}) and then multiplying by the
appropriate count rate to flux conversion factor
(\S\ref{cluster_count_rate}). Using the appropriate K-correction
(\S\ref{bkcorr}), we calculate the corresponding (0.5-2.0 keV)
luminosity. We compare this luminosity to the one obtained from
equation~(\ref{lum-eq}) using the input value of $r_c$. If the two
luminosities differ by more than 20\%, we recalculate $r_c$ and repeat
the whole procedure. We always start the iteration with $r_c=250$
kpc. Usually the process converges after only 1 or 2 adjustments to
$r_c$. After convergence, we define $L_{8\sigma}$ using the
appropriate bolometric correction (\S\ref{bkcorr}).

We note that, across all 144,000 calculations, the smallest $r_{50}$
value used to derive an $L_{8\sigma}$ value was $\simeq20 \arcsec$. To
confirm that clusters of this size, and larger, will be flagged as
extended sources, we have computed, using {\sc quicksim}, how the
radius enclosing half the flux in a model {\em XMM\/} point spread
function varies with off-axis angle. We find that the maximum size of
this radius is only $\simeq13 \arcsec$. Therefore, it follows that all
clusters detected at $>8\sigma$ will be flagged as extended sources,
i.e.~the completeness of the {\em XCS} should not suffer by the
imposition of an extent criterion.


It would be impractical to provide tables listing the results of all
144,000 calculations, so instead we give a few illustrative
examples. These examples are based on calculations in the first radial
bin ($\bar\theta=1'.5$) and in an $\Omega_0 = 1$, 
$\Omega_\Lambda = 0$ cosmology.
\begin{enumerate}

\item{The highest number of counts required inside $r_{50}$ for an 8
$\sigma$ detection was 255, arising for a cluster at redshift $z=2$
with a temperature $T=1$ keV, observed for 100 ks.  Such a cluster
would have a total count rate of 0.0051 count s$^{-1}$, a flux of 7.06
$\times 10^{-15}$ erg s$^{-1}$ cm$^{-2}$ and a luminosity of 7.48
$\times 10^{44}$ erg s$^{-1}$.}

\item{The lowest number of counts required inside $r_{50}$ for an 8
$\sigma$ detection was 77, arising for a cluster at redshift $z=0.35$
with a temperature of $T=12$ keV, observed for 5 ks. Such a cluster
would have a total count rate of 0.0308 count s$^{-1}$, a flux of 3.11
$\times 10^{-14}$ erg s$^{-1}$ cm$^{-2}$ and a luminosity of 0.16
$\times 10^{44}$ erg s$^{-1}$.}

\item{All clusters lying at $z\la1.5$ which are brighter than
$L_\star$ will be detected at $>8\sigma$ in a 5 ks pointing.  At
$z=1.5$ an $L_\star$ cluster has a flux of 4.24 $\times 10^{-14}$ erg
s$^{-1}$ cm$^{-2}$.  Here we assume that $L_\star$, the `knee' in the
Schechter function fit to the X-ray cluster luminosity function, has a
value of $4.8\times10^{44}$ erg s$^{-1}$: this is the average of the
values found by De Grandi et al.~(1999a) and Ebeling et
al.~(1998). Such a cluster has a temperature of 5.5 keV based on the
$L_{\rm X}-T$ relation of AF98, equation~(\ref{l_t_af98}).}

\item{In the average exposure time (22.3 ks, \S\ref{extdist}) of the
760 GTO pointings, it will be possible to detect clusters brighter
than $2.4\times10^{44}$ erg s$^{-1}$ (i.e.~$0.5\times L_{\star}$) out
to redshifts of $z\simeq1.6$. (Such a cluster has a temperature of 3.7
keV based on the $L_{\rm X}-T$ relation of AF98.)}

\end{enumerate}

\subsection{Calculation of Sensitivity Limits for Temperature Estimation}
\label{temp-est}

The EPIC--pn camera is able to estimate the energies of all 
incident photons, so it can perform
low--resolution spectroscopy as well as broad--band imaging. This
means that we are able to estimate temperatures for the clusters we
detect, provided the signal--to--noise ratio of the source spectrum is
sufficiently high. This is clearly a great advantage, since, for those
clusters for which it is possible, we shall not need to obtain
follow--up observations to determine their temperature (c.f. the {\em
ASCA\/} follow--up of {\em EMSS\/} clusters by Henry 1997). Cluster
temperature measurements are important for cosmological parameter
estimation, since $T$ is more readily related to cluster mass (in
terms of which theoretical predictions are made) than is (the more
easily measured) X--ray luminosity.

To assess the extent to which we will be able to take advantage of the
{\em XMM} spectral resolution, we have calculated the minimum
bolometric luminosity ($L_T$) that would yield a temperature estimate
for each of our 144,000 parameter combinations.  We stress that, in
most cases, determination of the redshift of the cluster will be
required before its temperature can be estimated. This remains true
even for spectra of high signal--to--noise, due to the degeneracy
between temperature and redshift in the spectral fitting when thermal
bremsstrahlung is the dominant emission process.  However, if there is
significant line emission in addition to the bremsstrahlung radiation
(which is especially true for low temperature, high metallicity
plasmas), it is sometimes possible to measure the redshifts from
emission features such as the 7 keV Fe Line.  Mushotzky (1994) and
others have shown that this technique works and we will certainly
apply it where possible to {\em XCS\/} data.  Alternatively, we might
also expect to be able to obtain crude redshift estimates using the
measured flux and extent: this method is cosmology--dependent, but is
still worthy of further investigation.

Given that the redshift will be known prior to the spectral fitting,
it will be possible to choose the metric aperture size most suitable
for temperature measurements for each cluster. To reflect this, we
allow the radius of apertures used in our temperature sensitivity
limit calculations to vary (with the constraint that it must never be
smaller than $r_{50}$), so as to include the maximum number of photons
but without being swamped by the background. (This is in contrast to
our detection sensitivity calculations, for which we always used
$r_{50}$, since the detection software will most likely only pick out
the central $\simeq50\%$ of the cluster flux, see \S\ref{inner50}).
We also set the additional criteria that the number of background
counts in the aperture must never exceed the number of cluster counts,
and that the cluster counts must never be less than 1000.  The
aperture sizes thus chosen varied from $r_{50}$ to $r_{89}$ and the
number of background counts inside these apertures varied from
$\simeq$250 to $\simeq$1500 (with an average value of $\simeq$600).

\begin{figure}
\centerline{\psfig{figure=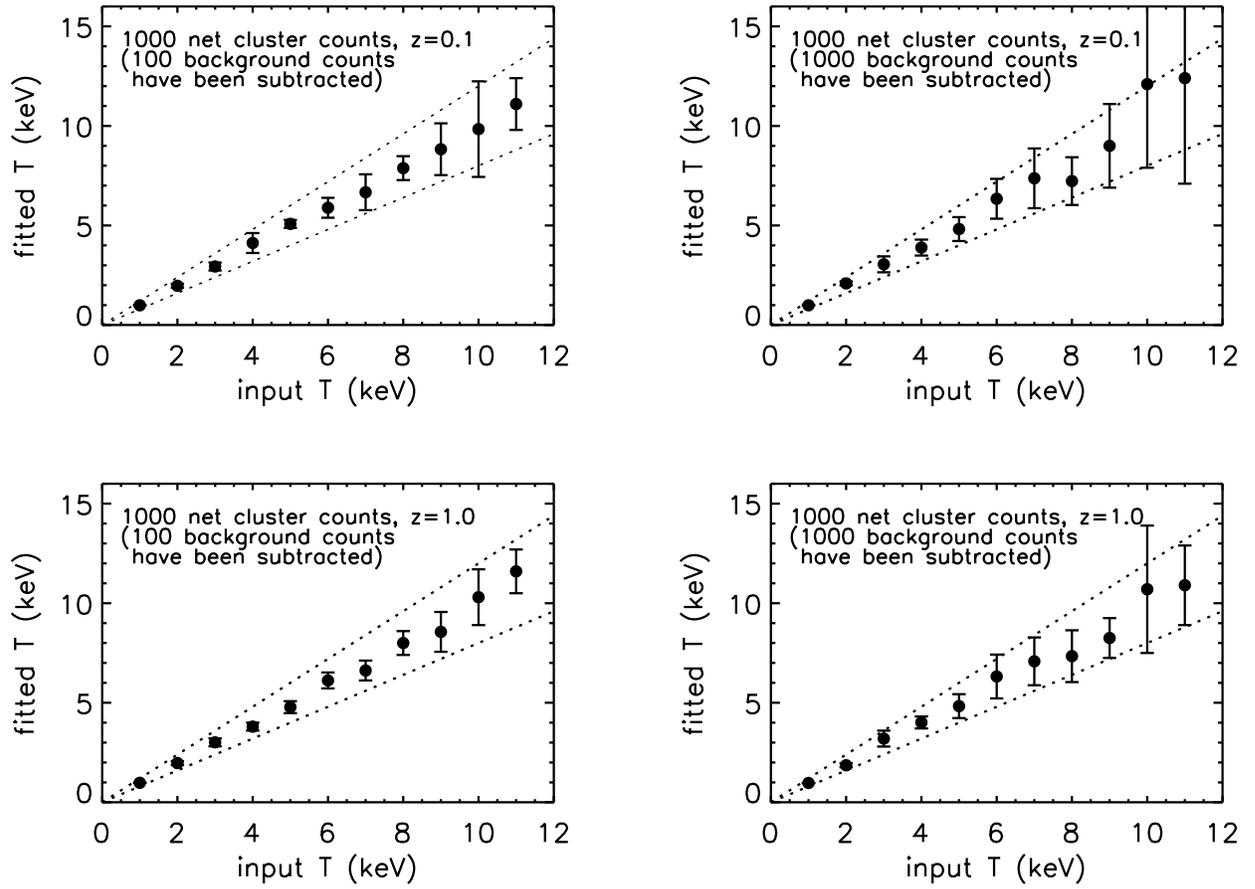}}
\caption{\label{temp-err} Fitted temperatures versus input temperatures 
for 4 different combinations of redshift and background contamination.
All spectra were created using {\tt fakeit} in {\sc xspec}. The dotted 
lines show input temperature plus (upper line) and minus (lower line) 20\%.}
\end{figure}

The accuracy to which temperatures can be estimated depends on three
factors: the cluster redshift, the cluster temperature, and the
signal--to--noise of the spectrum. We illustrate this via
Figure~\ref{temp-err}, which shows the input versus fitted temperatures
for representative values of the cluster redshift, cluster temperature
and background count rate. For this Figure, we created, and then
fitted, 20 fake spectra (with $n_H=4\times10^{20}$ cm$^{-2}$ and
$Z=0.3Z_{\odot}$) for each of the listed
temperature--redshift--background combinations using the {\sc xspec}
commands {\tt fakeit} and {\tt fit} respectively. The mean and
standard deviation of the 20 fits are plotted in Figure~\ref{temp-err}.
These fits were all performed on spectra containing 1000 counts,
because only about $1\%$ of the 144,000 calculations produced
background counts -- and hence cluster counts -- that exceeded 1000.
From the Figure, it is clear that a spectrum of 1000 counts will yield
temperatures estimates of varying accuracies, with the most accurate
values being derived for the lowest temperature systems.  The lowest
accuracy results will come from high-temperature clusters at low
redshift with high background count rates. We note that the accuracy
improves with redshift for high-temperature systems because the `knee'
in the thermal bremstrahllung spectrum moves to lower energies, where
{\em XMM} has more effective area.

Some examples of our temperature sensitivity limits are as follows
(assuming $\bar\theta=1'.5$, $\Omega_0 = 1$ and $\Omega_\Lambda = 0$).
Cool clusters ($T=2$ keV) will yield temperature measurements only out
to $z\simeq0.21$ in 5 ks exposures.  Even in a 100 ks exposure, the
maximum redshift for temperature determination for $T=2$ keV clusters
stretches only to $z\sim0.72$.  (Based on the AF98 $L_{\rm X}-T$
relation, we expect a $T=2$ keV cluster to have a luminosity of
$\simeq0.7\times 10^{44}$ erg s$^{-1}$.)  By contrast, hotter
clusters, which are brighter, yield temperature measurements to higher
redshifts. For example, temperatures will be measured for
$L_X>L_\star$ ($T>5.5$ keV) clusters to $z\sim0.6$ ($1.1$, $2.0$) in 5
(22.3, 100) ks exposures.

\subsection{Expected Catalogue Properties}
\label{results:catalogues}

\begin{figure}
\centerline{\psfig{figure=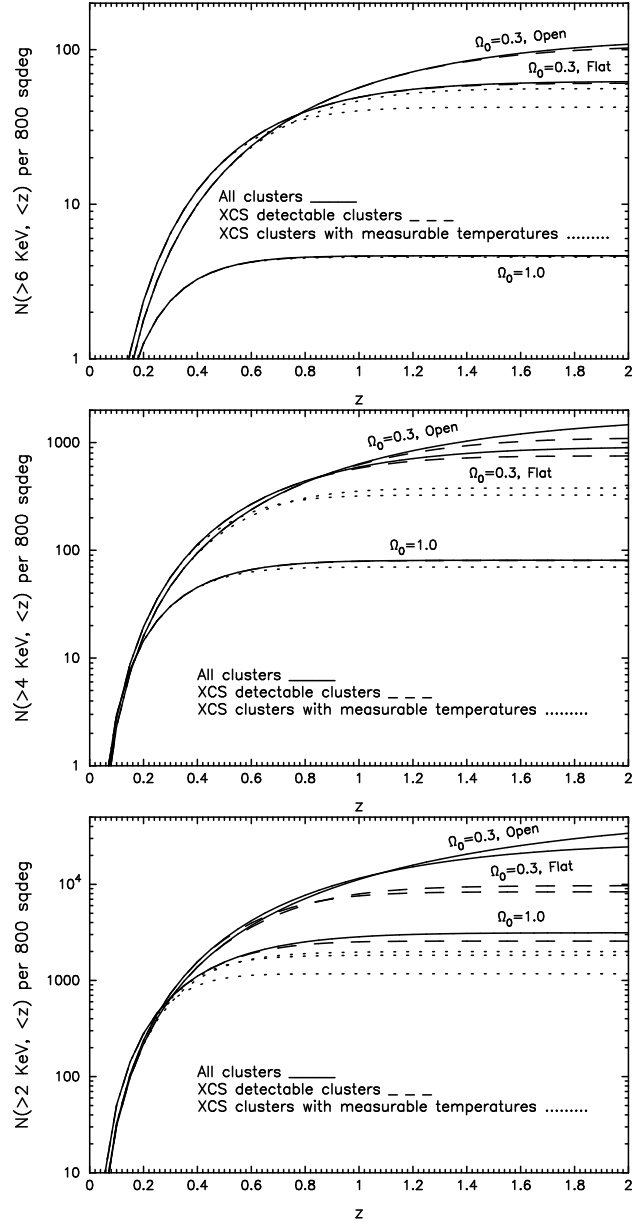,height=9.in}}
\caption{\label{numbltz} The cumulative redshift distribution,
$N(<z)$, of galaxy clusters per 800 square degrees with X-ray
temperature in excess of 6 KeV (upper panel), 4 keV (middle panel) and
2 keV (lower panel). The solid lines show the result one would obtain
if there was no limitation on the detectable flux. The dashed and
dotted lines show our predictions for the {\em XCS\/}, where the
dashed line represents the expected number of $>8\sigma$ detections and the
dotted line represents the expected number of clusters bright enough to allow 
temperature measurements.}

\end{figure}

\begin{figure}
\centerline{\psfig{figure=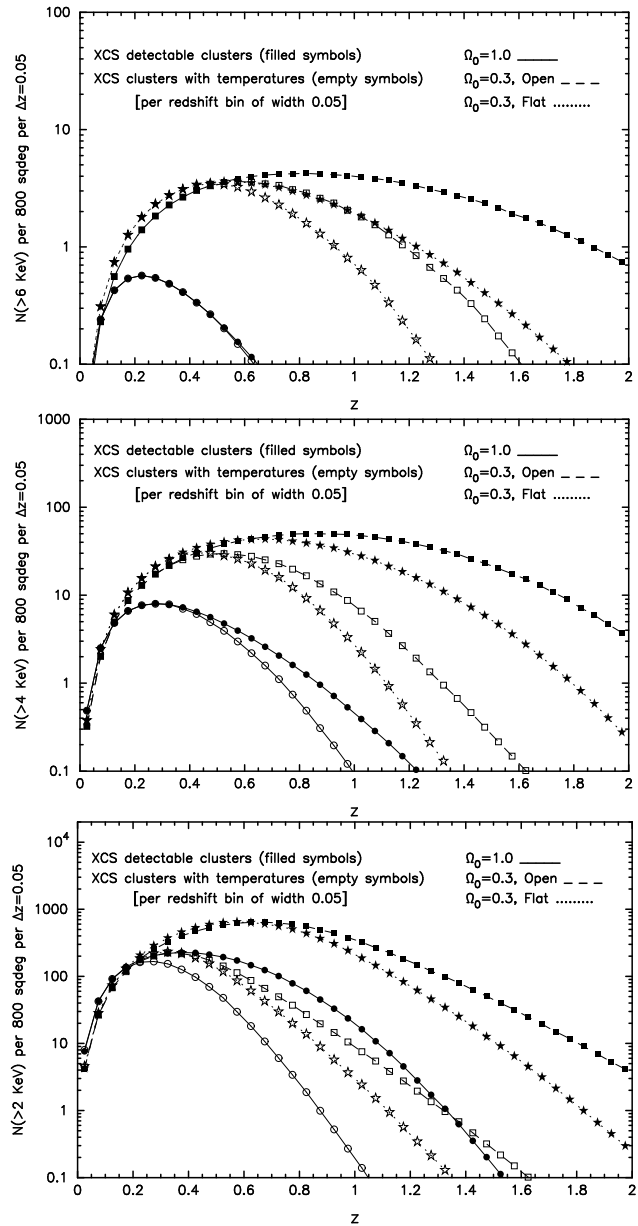,height=9.in}}
\caption{\label{numbdiffz} The predicted redshift distributions
(evaluated in bins of width $\Delta z=0.05$) corresponding to the
cumulative counts of Fig.~\ref{numbltz}. The filled and empty symbols
denote clusters detected and with temperatures estimated,
respectively, in the three cosmological models: Einstein - de Sitter 
(solid line), low density open (dashed) and low density flat (dotted).}
\end{figure}

In \S\ref{detectcalc} and \S\ref{temp-est} we computed the luminosity
threshold for cluster detection ($L_{8\sigma}$) and temperature
estimation ($L_T$), respectively, for 144,000 different combinations
of $\Omega_0, \Omega_\Lambda,T,z,\theta$ and $t$.  Combining these
luminosity thresholds with the results of \S\ref{ps_model} allows us
to estimate how many clusters will be included in our catalogue (and
for how many of them we can estimate a temperature). We note that,
when doing so, we assume that clusters are randomly located on the
sky, that the total areal coverage is 800 deg$^2$ (\S\ref{extdist})
and that the pointing exposure times are distributed according to
Table~\ref{exp-dist}.  The results of our catalogue predictions are
summarized in Table \ref{cluster-numbers} and Figs.~\ref{numbltz} and
\ref{numbdiffz}.

The {\em XCS\/} will not have a single, well defined flux limit,
because it will be made up of pointings with a wide dispersion of
exposure times (\S\ref{extdist}), and because -- in the {\em XMM} band
at least -- count rate to flux conversion factors are a complex
function of $z$ \& $T$ (\S\ref{cluster_count_rate}) and $\theta$
(\S\ref{vignetting}).  Despite this, we have been able to estimate an
effective flux-limit for the survey by comparing the numbers of
expected $z>0$ cluster detections (as listed on the first line of
Table~\ref{cluster-numbers}; 8300, 750, 61~etc.) with the $N(f)$
values in Fig.~\ref{fluxallz} (after appropriate scaling from $4\pi$
steradians to 800 deg$^2$). Doing so provides nine estimates of the
survey flux limit, all of which turn out to be close to
$1.5\times10^{-14}$ erg s$^{-1}$ cm$^{-2}$.  Repeating the procedure
for the $z<0.3$ and $z>1.0$ (by comparison with Figs.~\ref{fluxgt1}
and \ref{fluxlt0.3} respectively) also yields flux limits of $\simeq
1.5\times10^{-14}$ erg s$^{-1}$ cm$^{-2}$. We conclude, therefore,
that the effective {\em XCS} flux limit will be $\simeq
1.5\times10^{-14}$ erg s$^{-1}$ cm$^{-2}$, but stress that individual
pointings in the survey will have flux limits that may be higher or
lower than this value.

In \S\ref{ps_model} we noted that any survey that reaches a flux
limit of $\sim 1\times10^{-14}$ erg s$^{-1}$ cm$^{-2}$
will be able to detect almost all the $T>4$ keV clusters in its survey 
region, irrespective of redshift. This is supported by Fig.~\ref{numbltz},
which shows the integral redshift distributions, 
$N(z)$, predicted for the {\em XCS\/} in the three cosmologies we consider.
The total number of clusters (as predicted by Press--Schechter
theory) are depicted by solid curves, whereas the number of 
expected {\em XCS} detections are depicted by dashed curves. 
For $T>6$ keV clusters (top panel) the two curves are coincident
out to $z\simeq 1.4$, meaning that we can expect to detect all $T>6$ keV
clusters at $z<1.4$. From the middle panel, we can see that incompleteness
sets in earlier for the $T>4$ keV clusters, but even so we can expect to
detect almost all $T>4$ keV clusters out to at least $z\simeq1$. 
By contrast, we expect to be incomplete in terms of $T>2$ keV clusters by
$z\simeq 0.5$ and, by $z\simeq2$, we can expect to be detecting only 20\% 
of the $T>2$ keV clusters (if $\Omega_0=0.3$) in our survey region.

Also shown on Fig.~\ref{numbltz} are our predictions for the number of
clusters that we will detect with sufficient signal--to--noise to be able
to estimate temperatures (dotted curves). These numbers are also given
in parentheses in Table~\ref{cluster-numbers}. We will obtain temperatures
for all $T>6$, $T>4$ and $T>2$ keV clusters out to $z\simeq0.7$, 
$z\simeq0.5$ and $z\simeq0.3$ respectively.
To further illustrate the expected properties of the {\em XCS}, we
plot in Fig.~\ref{numbdiffz} the differential redshift distribution as
a function of cosmology and temperature. This figure shows how many
clusters will be detected in each $\Delta z=0.05$ bin when
$\Omega_0=1$ (solid lines, square symbols),
$\Omega_0=0.3,\Omega_\Lambda=0$ (dashed lines, star symbols) and
$\Omega_0=0.3,\Omega_\Lambda=0.7$ (dotted lines, circular symbols).
In order to differentiate between curves representing the number of
detections and curves representing the number of clusters with
temperature estimates, we have used solid and open symbols
respectively.

\section{Discussion and Conclusions}\label{discussion}

A cluster catalogue of the quality described in
\S\ref{results:catalogues} would have a great many uses. Here we
describe briefly a subset of these, to give a feel for the kind of
science which the {\em XCS\/} will make possible. We also discuss some
caveats relating to the methods used herein, and present our
conclusions.

\subsection{Science from the Catalogue}

The science that can be derived from the {\em XCS\/} can be loosely
divided into two categories: science which can be obtained directly
from the catalogue itself (for the most part assuming that follow--up
observations have provided cluster redshifts and enabled temperature
determination where possible), and those future projects which can
build upon the {\em XCS\/} data. We give a few examples of both sorts
of project here.

\subsubsection{Constraints on Cosmological Parameters}\label{cosmo-param}

The {\em XCS\/} cluster temperature and redshift distributions can be used as a
direct probe of the cosmological parameters $\Omega_0$ and
$\Omega_\Lambda$. The survey's size, redshift distribution and
selection criteria are ideally suited to this task. The {\em XCS} will
provide stringent constraints on $\Omega_0$ and has the potential to
offer the first constraint on $\Omega_\Lambda$ from cluster number
density evolution.  The power of the {\em XCS} to constrain these
parameters is clearly demonstrated in Fig.~\ref{numbltz}, from which
it is apparent that there is about an order of magnitude difference
between the number of high-temperature ($T>4$ keV) clusters in the
$\Omega_{0}=1$ case compared to either of the two $\Omega_{0}=0.3$
cases. At lower temperatures ($T>2$ keV) the differences between the
various models are less apparent, demonstrating that it is important
to concentrate on the high-temperature systems when attempting to
measure cosmological parameters.

It is beyond the scope of this paper to make detailed predictions of
the errors on $\Omega_0$ that would result from the {\em XCS}. To do
so would require a careful tracking of the theoretical uncertainties
in the number density predictions, especially those connected with the
amplitude of the power spectrum, and ideally would also take into
account the weak sensitivity of the predictions to quantities such as
the power spectrum shape. Further, the modeling of the observational
errors, and in particular the cosmic variance contribution -- which
assesses the extent to which the observations might be a statistical
fluke -- is a subtle business requiring detailed Monte--Carlo and
probably $N$-body simulations. The former can only be carried out in
detail once the true distribution of observing times, and the fraction
of usable pointings, is known. It may also prove necessary to model
evolution in the temperature--luminosity relation
(\S\ref{L-Tevolution}). (It is for these reasons, we are unable to add
error bars to the predictions in Figures~\ref{fluxallz},
\ref{fluxgt1}, \ref{fluxlt0.3},
\ref{numbltz} and \ref{numbdiffz}.)

In order to go beyond measurements of $\Omega_0$ and start to
constrain $\Omega_\Lambda$, one must study the $z>1$ population; from
Fig.~\ref{numbdiffz} we can see that there is little difference
between the cluster number density evolution predictions for the two
$\Omega_{0}=0.3$ cosmologies below $z=1$. But, for $z>1$, the number
density of galaxy clusters for $\Omega_{0}=0.3$ in open models is more
than twice that in flat models. Once all the $z>1$ clusters detected
have measured redshifts and temperatures, it should be possible to
constrain $\Omega_\Lambda$. However, in view of the modeling
uncertainties described above, it is premature to try and assess how
well that can be done, as this will only become apparent when the
actual data are available.

It is important to note that the cosmological constraints derived from
the {\em XCS} will be important even in the era of sensitive CMB
anisotropy experiments such as {\em Planck\/}, because the cluster
measurements can help to break degeneracies in cosmological parameters
inherent in CMB analyses.  Since the microwave anisotropy is expressed
in terms of angular scales on the sky, the cosmological parameters
$\Omega_0$ and $\Omega_\Lambda$ are only constrained in the
combination in which they arise in the angular diameter distance at
the redshift of the last scattering surface --- i.e.~such observations
can only constrain the Universe to lie somewhere along a line in the
($\Omega_0$,$\Omega_\Lambda$) parameter space, rather than at a single
point.  This degeneracy can be broken, in principle, in two ways. On
the very largest angular scales, it is weakly broken by the integrated
part of the Sachs--Wolfe effect, but these low-order multipoles suffer
a large `cosmic variance' which limits the accuracy with which they
can be estimated. On small angular scales, it is mildly broken by
gravitational lensing effects.  Both these effects are small, and it
is expected that the degeneracy will largely still be present even
after {\em Planck\/} has flown (Bond, Efstathiou \& Tegmark 1997). One
therefore requires other types of observations to break the
degeneracy. (See Bahcall et al.~1999 for an overview of how different
methods can be used in combination to constrain cosmological
parameters.) The sorts of measurements that can be used to break the
CMB $\Omega_0$--$\Omega_\Lambda$ degeneracy include cluster number
density evolution, large--scale structure analyses, and the
magnitude--redshift relation for Type Ia supernovae (Perlmutter et
al.~1999). Because each of these methods has intrinsic biases, it is
important to pursue all of them, to as high an accuracy as possible,
in order to derive a consistent model of the Universe.

\subsubsection{Evolution of X-ray Properties} \label{L-Tevolution}

The {\em XCS} will greatly improve our understanding of how cluster
properties, such as luminosity, temperature, metallicity, gas mass
fraction, core radius etc., evolve.  Importantly, it will allow the
luminosity--temperature ($L_{\rm X}-T$) relation to be measured in a
coherent fashion over a wide redshift range.  The quantification of
$L_{\rm X}-T$ evolution is crucial to our ambition to measure
cosmological parameters from cluster number densities.  From
Table~\ref{cluster-numbers}, we can see that the {\em XCS} will yield
temperature measurements for 1800 clusters with $T>2$ keV in the low-density 
flat cosmology.  These measurements will yield the most accurate
derivation of the $L_{\rm X}-T$ relation to date. Not only will the
derivation include a great many more clusters, but, for the first time, these
clusters will have been drawn from a single,  statistically complete
sample.

\subsubsection{Gravitational Lensing by {\em XCS\/} Clusters}

X--ray clusters magnify background galaxies via gravitational lensing.
This effect is well known at optical wavelengths (e.g~Luppino et
al.~1999), but, as shown by Smail, Ivison \& Blain (1997), it is
particularly exciting in the submillimeter. Here, the combination of
the lensing amplification and the positive K-correction in the
submillimeter (resulting from the sharp decline in the spectral energy
distribution of starburst galaxies longward of $\sim100\mu$m) means
such galaxies can be readily detected to extremely high redshift
($z>5$).  The follow--up of lensed galaxies around {\em XCS} clusters
with the coming generation of (sub)millimeter instruments, such as the
Atacama Large Millimeter Array, will provide an important insight into
the star formation history of the Universe. Lensing signals can also
be used to measure total masses and mass profiles of clusters in a
manner complementary to X-ray methods (e.g.~Squires et al.~1997).

\subsubsection{Sunyaev-Zel'dovich Follow-Up}\label{sz-targets}

The Sunyaev--Zel'dovich (SZ) effect (Sunyaev \& Zel'dovich 1972)
describes the inverse Compton scattering of cosmic microwave
background (CMB) photons to higher energies via interactions with hot
electrons in the ICM. Measurements of the SZ effect, in combination
with X-ray observations, provide a useful cosmological tool. They can
be used to constrain the value of the Hubble Parameter ($H_0$,
e.g. Birkinshaw 1999), the universal baryon fraction (e.g.~Grego
1998), cluster peculiar velocities (e.g.~Holzapfel et al.~1997) and
have the potential to place powerful constraints on the value of
$\Omega_0$ (e.g. Bartlett et al.~1998). Therefore, SZ follow-up of
{\em XCS} clusters will yield many important results, the most obvious
of which would be the measurement of $H_0$ as a function of
redshift. This $H_0(z)$ measurement would take advantage of the large
number of high $z$ clusters in the {\em XCS} and of the fact that the
SZ effect is redshift independent. The {\em XCS} also has the
potential to provide the required X-ray follow-up for blind SZ-surveys
(such as that proposed by Carlstrom et al.~2000). These blind surveys
hope to take advantage of the redshift independence of the SZ effect
in order to detect very distant clusters.

\subsubsection{Analysis of CMB Foregrounds} 

The limit to which the {\em MAP\/} and {\em Planck\/} satellites can
determine the power spectrum of CMB anisotropies on small scales is
likely to be set by the effectiveness of the foreground analyses. One
of the major sources of CMB foreground confusion is the SZ signal from
X-ray clusters of galaxies. Since the SZ effect is approximately
redshift independent, clusters at all distances will contribute to the
foreground signal. In a low-density cosmology, the mean SZ signal
comes from a broad range of redshifts out to $z\simeq2$ (da Silva et
al.~2000). The {\em XCS\/} will play a crucial role in the
understanding of this signal, because it will provide a statistical
description of the cluster population out to high redshifts.
Moreover, in the regions covered by the {\em XCS\/}, it will be
possible to mask out the signal from individual clusters from the CMB
maps.
 
\subsection{Limitations of our Calculations}\label{caveats}

As detailed in \S\ref{assumptions}, our simulations rest on a 
set of assumptions and simplifications, which were chosen because
they all seem reasonable given current knowledge. There are, however,
some limitations to the accuracy with which we can predict expectations
for the {\em XCS\/}, and we discuss some of them below.

\subsubsection{Contamination by Low-Mass Groups}

The application of an extent criterion will not be sufficient to
remove all the contamination in the cluster candidate list. Low-mass
groups (including ``fossil groups''; Ponman et al.~1994; 
Vikhlinin et al.~1999; Romer et al.~2000), and some very 
low-redshift galaxies, will also enter the list by
virtue of their extent.  Low mass (and hence low temperature) groups
are certainly interesting objects; they provide invaluable insight
into the processes of elliptical galaxy evolution, metal enrichment in
the intra-cluster medium, and the dynamics of extended dark halos
(Mulchaey \& Zabludoff 1998). However, they have a very limited role
to play in the derivation of $\Omega_0$ and $\Omega_\Lambda$ (because
of the increasing degeneracy between models as the temperature limit
is decreased, see Fig.~\ref{numbltz}). The Press--Schechter formalism
becomes unreliable below $T\simeq2$ keV, so we are not able to predict
the number of $T<2$ keV groups that will be detected by the {\em
XCS}. For typical values of the temperature and bolometric luminosity
of the intragroup medium ($T=1$ keV, $L_{\rm X}=10^{42}$ erg s$^{-1}$,
Mulchaey \& Zabludoff~1998), we have estimated the maximum redshift at
which a group would be detected by our survey to be $z\sim$ 0.05
(0.09, 0.17) in 5 (22.3, 100) ks. We are actively investigating ways
to flag potential $T<2$ keV objects using a combination of extent,
{\em XMM} spectra and cross--correlations with optical sky survey
data.

\subsubsection{Contamination by Point Source Emission}

We assume in \S\ref{results:catalogues} that every cluster we detect
at $>8\sigma$ will be included in the final {\em XCS\/} cluster
catalogue, but this may not always be the case. Some clusters might be
excluded if they are contaminated by point source emission, which
might originate from an active galaxy inside the cluster, a foreground
object, such as an M star, or a background object, such as a
quasar. Romer et al.~(2000) describe the case of an extended X--ray
source (RXJ0947.8) which was excluded from their cluster catalogue, on
the grounds of it being coincident with a $z=0.63$ quasar, despite
there being a spectroscopically  confirmed cluster at the same
position and redshift. A total of four clusters were rejected from the
Bright {\em SHARC\/} cluster catalogue because the quality of the {\em
ROSAT\/} data did not permit the cluster flux to be disentangled from
that of a contaminating point source. Romer et al.~(2000) claim that two of
these systems probably have sufficient, uncontaminated, flux to merit
inclusion in the Bright {\em SHARC\/} cluster catalogue, which
corresponds to an incompleteness of the whole catalogue at the $5\%$
level.  We expect the incompleteness level to be much lower than this
for the {\em XCS\/} since the improved spatial resolution of {\em
XMM\/} over {\em ROSAT\/} will significantly enhance our ability to
mask out point source contamination when measuring cluster fluxes. 
We expect therefore that the wrongful exclusion of clusters from the
{\em XCS\/} catalogue will occur only very rarely and have an
insignificant effect on our ability to use the {\em XCS\/} as a
cosmological tool.

\subsubsection{Effect of Assumptions about the Cluster Model}\label{cluster_limits}

Perhaps the greatest uncertainty in our calculations comes from the
simplified model of the distribution and state of the intracluster
medium we employed. Our use of the isothermal $\beta$--model was
justified in part by the results of Mohr et al.~(1999), who showed
that it well described the azimuthally--averaged properties of known
clusters. However, this work was carried out in the {\em ROSAT\/}
bandpass (0.5-2.0 keV), over which the emissivity of the X-ray gas is
almost insensitive to cluster temperature for $T\gtrsim2$ keV. And, as
emphasized recently by Ettori (1999), the assumption of a simple
isothermal $\beta$-model will lead to significant errors when a
cluster with a significant temperature gradient is observed in a broad
band bracketing the energy corresponding to its mean
temperature. Evidence to suggest that the cluster gas is not
isothermal comes from spatially resolved cluster temperature maps
(e.g. Markevitch et al. 1998) and from the so-called ``$\beta$
discrepancy'' (e.g. Sarazin 1988; Bahcall \& Lubin 1993), which
describes the fact that fitted values of $\beta$ are not consistent
with the values expected from the combination of cluster temperatures
with galaxy velocity dispersions.

We have also ignored the effect of cooling flows in the cluster core. 
A significant fraction of relaxed clusters have regions of cool, dense gas
in their cores (e.g. AF98) and, as Ettori (2000) has pointed out, a
modified version of equation~\ref{beta-eq} would be more
appropriate to describe such clusters.

More fundamentally, it is possible, of course, that the clusters we
detect at high redshift will not be virialised systems. Clusters in
the process of formation may have significant non--thermal components
to their X--ray luminosities, for example from shocks resulting from
subcluster merging.  A classic example of such a cluster is RX J0152.7
(Romer et al.~2000; Ebeling et al.~2000), which has a high total
luminosity ($8.26 \times 10^{44}$ erg s$^{-1}$, $z=0.83$) 
but is made up of at least 2
components.  It is not possible to predict, at this stage, what the
net effect of unvirialised systems will be on the properties of the
{\em XCS\/}, but the spatial and spectral resolution of EPIC should
help us to recognize such systems.  The {\em XCS\/} may even show that
clusters are {\em not} suitable as cosmological probes above a certain
redshift: indeed, perhaps effects such as these lie behind the
detection to date of a (possibly) surprisingly large number of massive
clusters at high redshift (Luppino \& Gioia 1995; Donahue 1996;
Luppino \& Kaiser 1997; Donahue et al.~1998; Eke et al.~1998), which
has been claimed to be troublesome for conventional models of
structure formation.

\subsection{Conclusions}

We have predicted the expected properties of a serendipitous cluster
survey based on archival {\em XMM\/} pointing data. We have done this
using simulations which combine a theoretical model of the properties
of the cluster population, as a function of cosmology, with a detailed
description of the characteristics of the EPIC camera, and a generic
model for cluster surface brightness profiles. We have shown that the
catalogue that would result from such a survey will surpass existing
catalogues of high--redshift ($z>0.3$) clusters in both size, quality
and redshift coverage, while, at low redshifts ($z<0.3$) the catalogue
will yield many more cluster temperature measurements than have ever
been measured before.

It is clear that, while the methods presented here may be adequate to
yield reasonably realistic predictions for what we can expect to get
from the {\em XCS\/}, the actual analysis of the data from the survey
will require a more sophisticated approach, informed by detailed
physical models resulting from pointed observations of individual
clusters made by {\em Chandra\/} and {\em XMM\/} itself.  This must,
however, be tempered by the requirement that the final set of cluster
selection criteria be readily modeled by Monte--Carlo methods: the
{\em XCS\/} will have its greatest impact in statistical analyses, so
it must be constructed in such a way that its selection function can
be well understood.

\section*{Acknowledgments}

We thank Mike Watson, Gordon Stewart, Julian Osborne, Bob Nichol,
Chris Collins for a series of useful discussion. We also thank
Monique Arnaud for her assistance with the cosmic background
spectrum. AKR acknowledges financial support from CMU physics
department and NASA grant NAG5-7926. PTPV was supported by 
the PRAXIS XXI program of FCT (Portugal) and RGM acknowledges support
from PPARC. Finally we thank our ApJ editor (John Huchra) and an 
anonymous referee for their helpful comments.



\clearpage

\clearpage
\begin{table*}
\caption{Coefficients of the quadratic fits to K-corrections for
Raymond--Smith spectra.}
\label{k-corr-tab}
\begin{center}
\begin{tabular}{|cccc|}\hline
\multicolumn{4}{|c|}{\bf 0.5-2.0 keV}\\ 
Temperature (keV)   & $c$ & $b$ & $a$ \\ \hline
       1&      1.00102&    -0.661006&    0.520801\\
       2&      0.98006&    -0.111214&    0.074738\\
       3&      1.07562&    -0.172770&    0.045052\\
       4&      1.14437&    -0.238201&    0.041994\\
       5&      1.19418&    -0.290640&    0.044047\\
       6&      1.23139&    -0.331765&    0.047111\\
       7&      1.25583&    -0.359344&    0.049642\\
       8&      1.27631&    -0.383054&    0.052145\\
       9&      1.29229&    -0.401918&    0.054310\\
      10&      1.30579&    -0.418036&    0.056253\\
      11&      1.31657&    -0.430874&    0.057859\\
      12&      1.32727&    -0.443794&    0.059521\\ \hline
\multicolumn{4}{|c|}{\bf 0.5-10.0 keV}\\ 
Temperature (keV)   & $c$ & $b$ & $a$ \\ \hline
       1&      0.98976&    -0.693811&    0.589274\\
       2&      0.74183&    ~0.168149&    0.076622\\
       3&      0.78837&    ~0.169155&    0.029814\\
       4&      0.85558&    ~0.106416&    0.021304\\
       5&      0.91672&     0.040785&    0.021762\\
       6&      0.96909&    -0.018076&    0.024729\\
       7&      1.01385&    -0.068723&    0.028104\\
       8&      1.05094&    -0.111368&    0.031331\\
       9&      1.08256&    -0.147891&    0.034250\\
      10&      1.10903&    -0.178722&    0.036790\\
      11&      1.13138&    -0.204740&    0.038903\\
      12&      1.15220&    -0.229466&    0.041146\\ \hline 
\end{tabular}
\end{center}
\end{table*}
\begin{table*}
\caption{Conversion factors between observed and pseudo-bolometric
luminosity for Raymond--Smith spectra.}
\label{b-corr-tab}
\begin{center}
\begin{tabular}{|ccc|}\hline
Temperature (keV)   & 0.5-10.0 keV & 0.5-2.0 keV \\ \hline
       1&      1.75287            &2.00900\\
       2&      1.48716            &2.37600\\
       3&      1.36920            &2.69564\\
       4&      1.33470            &3.04131\\
       5&      1.33501            &3.38720\\
       6&      1.35271            &3.72926\\
       7&      1.38674       	  &4.04849\\
       8&      1.42370            &4.36129\\
       9&      1.46514            &4.66436\\
      10&      1.50744            &4.95957\\
      11&      1.55192            &5.24149\\
      12&      1.59530            &5.53199\\ \hline
\end{tabular}
\end{center}
\end{table*}

\begin{table*}
\caption{Distribution of {\em XMM} 
exposure times in the GTO and AO1 observing cycles.}
\footnotesize{
\label{exp-dist}
\begin{tabular}{lccccccccccc}\hline
time (ks)  &5-10 &10-15&15-20&20-25&25-30&30-35&35-40&40-45&
45-50&50-55&55-100\\ 
percentage (GTO) &23.16&18.03&7.63 &17.50&4.87 &6.58&1.45 &7.24 &
1.58 &6.58 &5.39 \\ 
percentage (AO1) &23.79&18.97&4.41 &18.81&5.56 &9.08&1.31 &4.66 &
0.57 &7.23 &5.56 \\ \hline
\end{tabular}
}
\end{table*}

\begin{table*}
\caption{The expected number of clusters detected in an XMM serendipitous
survey covering 800 deg$^2$, for
three different cosmological models. The main numbers are for detections, 
while 
the numbers in parentheses are detections with sufficient flux to yield 
temperatures. }
\label{cluster-numbers}
\begin{center}
\begin{tabular}{|c|ccc|}
\hline
& \multicolumn{3}{|c|}{\bf $\Omega_0=0.3, \Omega_\Lambda=0.7$}\\
& $T>2$ & $T>4$ & $T>6$ \\ \hline
$z>0$& 8300 (1800)& 750 (320)& 61 (42)\\
$z>0.3$& 7600 (1200)& 700 (270)& 54 (36)\\ 
$z>1$& 750 (6)& 170 (6)& 12 (2)\\ \hline
& \multicolumn{3}{|c|}{\bf $\Omega_0=1.0, \Omega_\Lambda=0.0$}\\
& $T>2$ & $T>4$ & $T>6$ \\ \hline
$z>0$& 2600 (1200)& 80 (70)& 5 (5)\\ 
$z>0.3$& 1900 (570)& 50 (40)& 2 (2)\\ 
$z>1$& 46 (0)& 1 (0)& 0 (0)\\ \hline
& \multicolumn{3}{|c|}{\bf $\Omega_0=0.3, \Omega_\Lambda=0.0$} \\
& $T>2$ & $T>4$ & $T>6$ \\ \hline
$z>0$&9700 (2000)& 1100 (380)& 110 (56)\\ 
$z>0.3$&9000 (1400)& 1100 (330)& 100 (51)\\ 
$z>1$& 1700 (26)& 480 (24)&   50 (9)\\ \hline 
\end{tabular}
\end{center}
\end{table*}

\end{document}